\documentclass[pdflatex,sn-mathphys-num]{sn-jnl}

\usepackage{graphicx}%
\usepackage{multirow}%
\usepackage{amsmath,amssymb,amsfonts}%
\usepackage{amsthm}%
\usepackage{mathrsfs}%
\usepackage[title]{appendix}%
\usepackage{xcolor}%
\usepackage{textcomp}%
\usepackage{manyfoot}%
\usepackage{booktabs}%
\usepackage{algorithm}%
\usepackage{algorithmicx}%
\usepackage{algpseudocode}%
\usepackage{listings}%
\usepackage{svg}
\usepackage{natbib}
\usepackage{amsmath}
\usepackage[left]{lineno}

\newcommand{\be}{\begin{equation}}
\newcommand{\ee}{\end{equation}}
\newcommand{\bea}{\begin{eqnarray}}
\newcommand{\eea}{\end{eqnarray}}

\theoremstyle{thmstyleone}%
\theoremstyle{thmstyletwo}%

\theoremstyle{thmstylethree}%

\begin{document}


\title[PNC Paper]{Redefining the limits of real-time noise cancellation in optical fiber links}

\author*[1]{\fnm{Charles A.} \sur{McLemore}}\email{charles.mclemore@nist.gov}

\author[2]{\fnm{Marco}\sur{Pomponio}}

\author[1,3]{\fnm{Takuma}\sur{Nakamura}}

\author[1,4]{\fnm{Yifan}\sur{Liu}}

\author[1]{\fnm{Nazanin}\sur{Hoghooghi}}

\author*[5]{\fnm{Antonio}\sur{Mecozzi}}\email{antonio.mecozzi@univaq.it}

\author*[1,6]{\fnm{Franklyn} \sur{Quinlan}}\email{franklyn.quinlan@nist.gov}

\affil[1]{\orgdiv{Time and Frequency Division}, \orgname{National Institute of Standards and Technology}, \orgaddress{\street{325 Broadway}, \city{Boulder}, \state{CO} \postcode{80305}, \country{USA}}}

\affil[2]{\orgname{Temporis Solutio LLC}, \orgaddress{\city{Boulder}, \state{CO}, \country{USA}}}

\affil[3]{\orgdiv{Department of Electrical Engineering}, \orgname{University of Colorado Denver}, \orgaddress{\city{Denver}, \postcode{80217}, \state{CO}, \country{USA}}}

\affil[4]{\orgdiv{Department of Physics}, \orgname{University of Colorado Boulder}, \orgaddress{\city{Boulder}, \postcode{80309}, \state{CO}, \country{USA}}}

\affil[5]{\orgname{University of L’Aquila}, \orgaddress{\city{L’Aquila}, \postcode{67100}, \country{Italy}}}

\affil[6]{\orgdiv{Department of Electrical, Computer and Energy Engineering}, \orgname{University of Colorado Boulder}, \orgaddress{\city{Boulder}, \postcode{80309}, \state{CO}, \country{USA}}}

\abstract{A broad and growing array of applications rely on the faithful transmission of ultrastable optical signals over noisy paths, requiring cancellation of environmentally induced noise. A generally accepted limit constrains how well the path length noise can be suppressed in real time. Here, we show that this standard limit is not fundamental and can be improved upon. By considering the temporal correlations between the round-trip and one-way signals, we develop a new framework for optimizing the noise cancellation feedback signal for any spatial distribution of noise along the signal path. We use this framework to surpass the standard limit in two sets of experiments. First, we demonstrate noise cancellation in a deployed urban optical fiber, where we achieve noise suppression approximately 6~dB beyond the standard limit. Then, in a reconfigurable lab-based fiber-optic testbed, we show that, for certain spatial distributions of noise, suppression of well over 10~dB beyond the standard limit is readily achievable. With the use of digital signal processing to generate the correction signal, our new technique requires no new electro-optic hardware relative to the field-standard noise cancellation scheme. This will allow for widespread adoption of these improved limits in existing systems, with applications in optical clock distribution, optical clock comparisons for fundamental physics and geodesy, and quantum networking.}


\maketitle


Transmission of optical signals through deployed fibers is a cornerstone of modern technology, quietly enabling nearly every facet of modern life. Remarkably, the same technology that gives the world high-speed internet can also be leveraged to detect earthquakes~\cite{marra2018ultrastable,marra2022optical,lindsey2021fiber}, measure the shape and structure of the earth~\cite{mehlstaubler2018atomic,cecilia2015fiber,grotti2018geodesy}, and shed light on the great mysteries of the universe, from dark matter~\cite{manley2023searching} to changes in the fine structure constant~\cite{roberts2020search}. However, while deployed fibers provide high-speed and low-loss optical transmission, they also couple strongly to environmental fluctuations, often causing difficulty for precision applications. Vibrations, temperature variations, and other effects can change the path length that transmitted light experiences, distorting optical signals. For applications such as optical clock comparisons~\cite{collaboration2021frequency,rosenband2008frequency,nemitz2016frequency,dorscher2021optical,kim2023improved,lindvall2025coordinated}, ultrastable time and frequency transfer~\cite{caldwell2025high,clivati2020common,li2024phase,Hoghooghi:25,schioppo2022comparing,cantin2021accurate,clivati2022coherent,cizek2022coherent,bercy2014}, and quantum networking~\cite{nichol2022elementary,nakamura2025sub,bersin2024development,kucera2024demonstration,chung2021illinois,duan2001long,nardelli2025phase,liu2024creation,stolk2024metropolitan}, all of which rely on the faithful transmission of low-noise signals, these environmental distortions are a major impediment.

To deliver ultrastable optical signals from one location to another, real-time measurement and active compensation of path length fluctuations are needed. Following the seminal work of Refs.~\cite{Ma:94,bergquist1994laser}, this is typically accomplished by constructing a Michelson-type interferometer with the fiber path as one arm, thereby measuring the total phase noise accumulated by light that has double-passed the fiber. A correction signal is then generated that is proportional to half of the round-trip phase noise and is applied to the outgoing light field to compensate for the phase fluctuations it will experience along the way. While this technique has proven highly effective at suppressing fiber noise, it cannot perfectly cancel all environmental fluctuations. For the typical scheme described here, there is a widely accepted limit on how well fiber noise can be suppressed in real time, which we will refer to as the `standard limit'~\cite{newbury2007coherent,Williams:08}. For spatially uniform and uncorrelated noise, in the low Fourier frequency limit, this is 

\begin{equation}
    S_\mathrm{remote}(\omega)=\frac{1}{3}(\omega \tau)^{2} S_\mathrm{fiber}(\omega),
    \label{eqstandardlimit}
\end{equation}

\noindent where $\omega$ is the angular Fourier (or noise) frequency, $\tau$ is the one-way propagation delay through the fiber, $S_\mathrm{remote}(\omega)$ is the phase noise power spectral density (PSD) of the delivered light, and $S_\mathrm{fiber}(\omega)$ is the free-running, one-way phase noise PSD of the fiber.

It is worth emphasizing that this standard limit dictates how well the excess fiber noise on a delivered signal can be suppressed in real time. Some applications of fiber links are able to utilize post-processing, the collocation of the two ends of the fiber, or the use of stable laser light at the remote end~\cite{Stefani:15,calosso2015doppler,jabir2025enabling}. In these cases, with stabilized lasers at both ends of the fiber, the noise can be improved by 6 dB. However, the important cases of ultrastable frequency dissemination and certain phase-stable quantum networks~\cite{duan2001long,nardelli2025phase} require real-time, one-way noise cancellation. To the best of our knowledge, the standard limit has not yet been surpassed under these conditions.
Moreover, no existing real-time ultrastable laser delivery schemes can be tailored to situations where there are large, localized noise fluctuations, such as in links with aerial fiber sections, where greater noise suppression would be of tremendous value.

In this paper, we introduce an adaptable technique to surpass the standard limit in real time. We show that by exploiting the temporal correlations between the round-trip noise and the one-way fiber noise, it is possible to optimize active noise cancellation for an arbitrary fiber noise distribution. In the case of spatially uniform, uncorrelated noise, this results in a 6~dB improvement in phase noise over the standard limit, implying a factor of two improvement in Allan deviation. For more localized fiber noise distributions (where most of the noise is in one section of the fiber link), we show that noise cancellation of more than 10~dB beyond the standard limit is readily achievable. We demonstrate these new limits using a digital servo loop to implement our scheme in two optical fiber systems: a 5.5~km deployed urban fiber and an 8.8~km lab-based fiber with an adjustable noise distribution. By introducing a new framework for understanding optimal fiber noise suppression, this work enables active cancellation of optical fiber paths with arbitrary noise distributions beyond what has been viewed as possible. Furthermore, since this technique does not require significantly different hardware than the traditional setup, these noise improvements should be readily attainable with only minor changes to existing systems.

\section*{Results}\label{sec2}

\subsection*{Theoretical Framework}

All real-time ultrastable laser delivery techniques have the same basic structure: light that has double-passed the fiber path and returned to the local end is used to estimate the phase noise that light will experience when traversing a single pass. This estimate is used to generate a correction signal, which is applied to an acousto-optic modulator (AOM) or other actuator, to pre-compensate the departing signal for the one-way noise it will experience on its way to the remote end. The fundamental challenge is then to determine how best to use the double-passed light to estimate the one-way noise. To first order, the total phase noise accumulated over the round trip should be halved. (This is the standard practice.) But we show that full optimization requires examination of the temporal correlation between the round-trip and one-way signals. Specifically, the correction signal should be proportional to a \textit{temporally shifted} version of the round-trip signal, and the optimal temporal shift can be determined by the correlation function. In this section, we provide a pictorial derivation of the correlation function and discuss how it is used to provide improved cancellation of the one-way noise. A rigorous derivation is provided in Methods.

\begin{figure}[!htb]
    \centering
    \makebox[\textwidth][c]{\includegraphics[width=0.7\paperwidth]{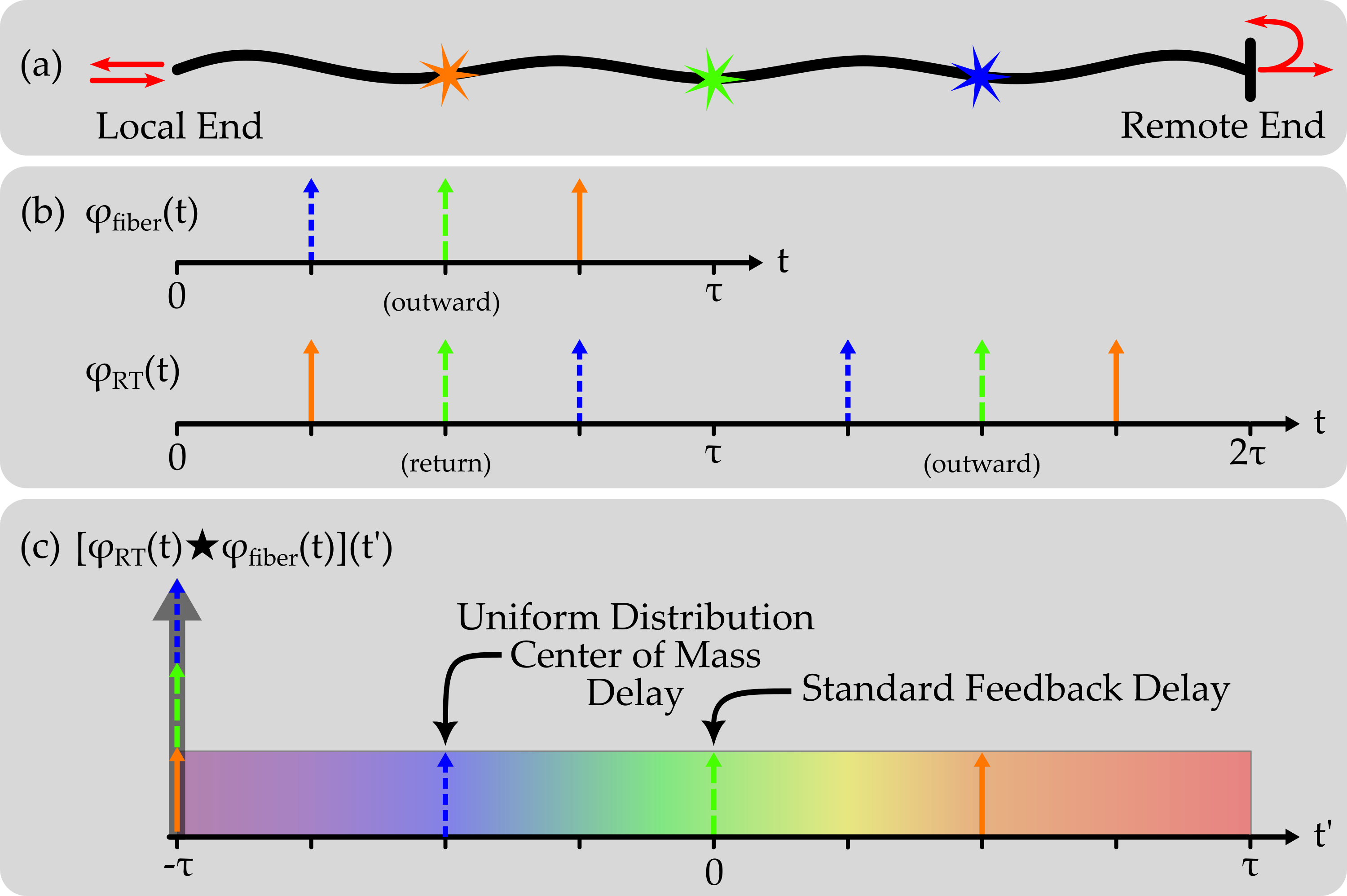}}
    \caption{\textbf{The `center of mass' picture of fiber noise.} (a)~A simplified distribution of noise along a fiber path with three localized, simultaneous `events' along an otherwise quiet fiber. (b)~Phase noise of the one-way outgoing light, $\varphi_\mathrm{fiber}(t)$, and the round-trip light, $\varphi_\mathrm{RT}(t)$. The latter signal has the three events encoded twice each: once on light on its outward journey and once on returning light. (c)~Cross-correlation of the round-trip signal and the one-way light. Each of the three events appears twice in the cross-correlation. The first instance comes from the overlap of the one-way light with the return journey of the round-trip light, resulting in a spread in the cross-correlation from $-\tau$ to $\tau$. The second instance comes from the overlap between the one-way light and the outward path of the round-trip light; these overlaps occur at a delay of $\tau$ for all three events, resulting in a strong peak at $t'=-\tau$. Rainbow shading extrapolates from the `three-event' case above to a uniform noise distribution along the fiber. The center of mass of the uniform distribution cross-correlation function is located at $t'=-\tau/2$, requiring a compensating temporal advance relative to the delay imposed by the standard feedback scheme.}
    \label{fig_intuitive}
\end{figure}

We begin with the simplified picture of Fig.~\ref{fig_intuitive}a, which depicts a fiber path with three distinct noise events that occur simultaneously at different locations along the fiber. 
Light is sent through this fiber to the remote end, where some of the light is delivered and some is reflected back to the local end. Noise from each of the three events is carried on the outgoing light and is seen once at the remote end of the fiber, in proper order. The local end of the fiber will experience each event twice: first in reverse order on light that was already returning at the time of the events, and then in proper order on light that was outgoing at $t=0$ and subsequently reflected back at the remote end.
The phase noise from these events is represented in Fig.~\ref{fig_intuitive}b, where the arrows indicate the time of arrival of the fluctuations at either the local or remote end, and are color-coded to correspond to the three events. Thus, we see that, in addition to the total round-trip noise amplitude being double that of the one-way noise, the round-trip noise is temporally delayed. Since each event appears twice on the round-trip noise with different delays, it is the center of mass that is used to describe its effective delay. For example, when the noise is at the center of the fiber (green `event'), the relative delay between the one-way noise (at $t=\tau/2$) and the round-trip center of mass (at $t=\tau$) is $\tau/2$. Since the round-trip center of mass is always at $\tau$ and the one-way noise varies from 0 to $\tau$, the relative delay between the two signals can vary between 0 and $\tau$, depending on whether the noise is closer to the local or remote end of the fiber. The optimized correction signal will counteract this delay by applying a corresponding temporal advance.

To generalize the delay for arbitrary noise distributions, we next consider the cross-correlation between the one-way and the round-trip noise. The cross-correlation function is plotted in Fig.~\ref{fig_intuitive}c, created from the plots of Fig.~\ref{fig_intuitive}b by simply sliding the round-trip signal by a shift of $t'$ relative to the one-way signal and marking the delays where same-colored arrows overlap. In this Figure, the delay variable $t’$ is negative when round-trip noise lags the one-way noise. The correlation function consists of a spike at $t' = -\tau$ as well as a spread from $-\tau$ to $+\tau$. Importantly, the area of the function for the spread from $-\tau$ to $+\tau$ is always the same as that of the spike at $t' = -\tau$. For noise at the center of the fiber, the center of mass of the correlation function is therefore at $t' = -\tau/2$, as expected. For noise localized at the local end, the correlation function is centered near $t' = 0$, whereas it is centered near $t' = -\tau$ for noise localized at the remote end. These observations can be generalized: the \textit{spatial} center of mass along the fiber corresponds to the \textit{temporal} center of mass of the correlation function. That is, once the center of mass of the noise along the fiber is known, so too is the temporal shift that will optimally cancel the noise.

If we parameterize the relative temporal shift between the round-trip and one-way signals as a fraction $\kappa$ of the one-way transit time $\tau$, we can write the phase noise of the (corrected) delivered light as

\begin{equation}
    \varphi_\mathrm{remote}(t)= \varphi_\mathrm{fiber}(t)-\frac{1}{2} \varphi_\mathrm{RT}(t+\kappa \tau),
    \label{eqnewscheme}
\end{equation}


\noindent where positive $\kappa$ corresponds to a temporal advance of the round-trip light. Here, $\varphi_\mathrm{fiber}(t)$ and $\varphi_\mathrm{RT}(t)$ are the one-way and round-trip phase noise signals depicted in Fig~\ref{fig_intuitive}b. In the case of a uniform, spatially uncorrelated fiber noise distribution, this leads to a maximum suppression of the one-way noise of approximately

\begin{equation}
   S_\mathrm{remote}(\omega)=(\kappa^2-\kappa+\frac{1}{3})(\omega \tau)^2 S_\mathrm{fiber}(\omega)
   \label{equniformlimit}
\end{equation}


\noindent in the low frequency limit (see Methods). This function is optimized for $\kappa=1/2$, which corresponds to the expected temporal advance of $\tau/2$. Notably, with $\kappa=0$ we recover the standard limit (equation~(\ref{eqstandardlimit})), which is four times higher than the optimal $\kappa=1/2$ case.


Of course, a true temporal advance of the round-trip noise in real time violates causality, and can therefore only be applied in post-processing~\cite{Stefani:15}. However, an \textit{effective} real-time advance can be achieved for noise frequencies where $\omega \tau \ll 1$. This is achieved by recognizing that the round-trip signal received at the local end, in addition to containing the round-trip noise, also contains the correction signal applied at a time $2\tau$ prior. This, in effect, stores the round-trip noise from an earlier time. It is by properly weighting and adding delayed copies of the correction signal that we may achieve the desired effective temporal shift. (This process assumes the noise does not change much after a round trip, imposing the $\omega \tau \ll 1$ condition.) Notably, the standard scheme also uses the delayed correction signal to provide a phase shift (corresponding to an effective temporal advance of $\tau$) at the local end of the fiber, such that there is zero temporal shift by the time light reaches the remote end. From this point of view, the standard noise-cancellation architecture is a special case of this more general noise-correction scheme.

\subsection*{Experimental Implementation}

\begin{figure}[!htb]
    \centering
    \makebox[\textwidth][c]{\includegraphics[width=0.75\paperwidth]{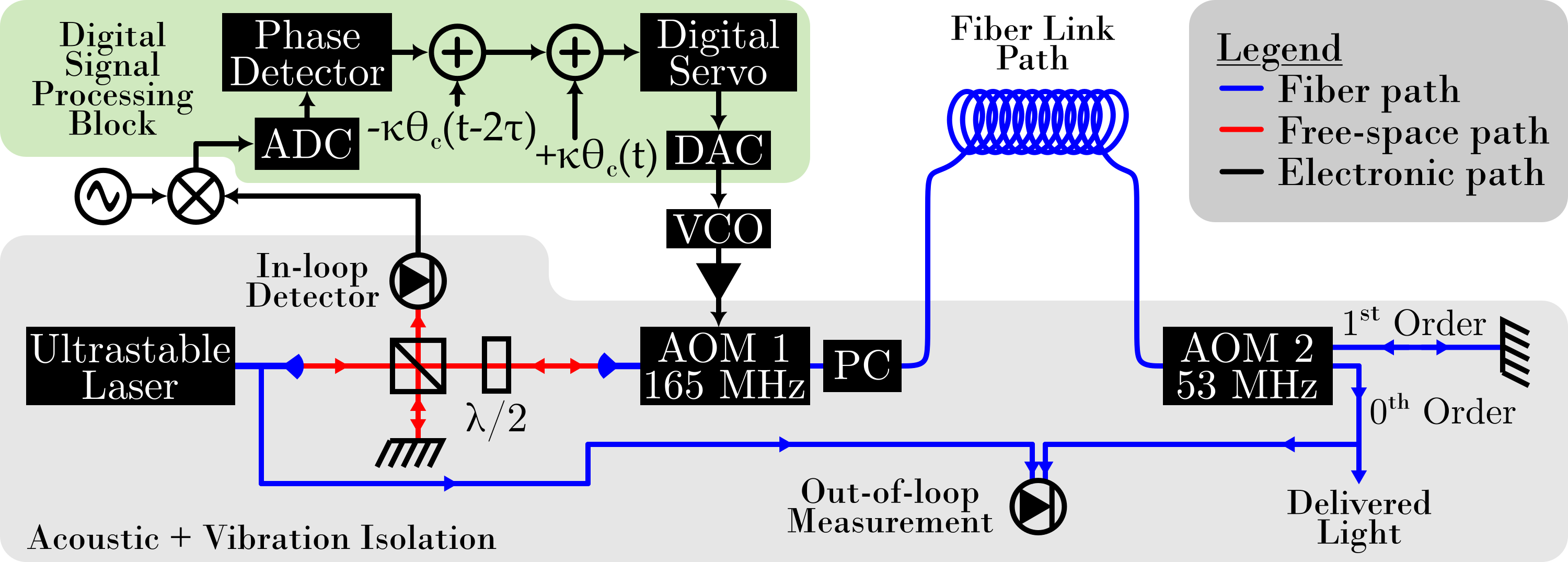}}%
    \caption{\textbf{Setup for optimized fiber noise cancellation.} The round-trip phase noise is detected with a Michelson interferometer. The effective temporal advance is then applied digitally by adding appropriately weighted and delayed copies of the correction signal $\theta_c(t)$, which is stored in memory for at least $2\tau$. This temporally advanced signal is then filtered, amplified, and applied as a correction signal to the feedback acousto-optic modulator (AOM). ADC, analog-to-digital converter; DAC, digital-to-analog converter; VCO, voltage-controlled oscillator; PC, polarization controller.}
    \label{digitalsetupfig}
\end{figure}

To implement this new feedback scheme, we need only slightly modify the standard layout for fiber noise cancellation. As shown in Fig.~\ref{digitalsetupfig}, the basic configuration is that of a Michelson interferometer.
An ultrastable laser is split into a short ($_{^{^{\sim}}}1$~cm), free-space reference arm and a long, fiber-optic delivery arm containing the fiber link path; upon reflection, the two beams are combined on an in-loop detector to generate a heterodyne beat note for noise cancellation. In the long arm of the interferometer, an AOM at the local end applies the noise cancellation to both the outgoing and return light. The first-order output of an AOM at the remote end is used simply to identify light that has traveled the full length of the fiber with a frequency shift, thereby minimizing the impact of spurious reflections. Part of the light at the remote end is reflected back through the fiber link path, while the rest is used for ultrastable laser delivery. In our experiments, the local and remote ends are collocated, allowing for an out-of-loop heterodyne between a portion of this delivered light and light directly from the ultrastable laser to assess the performance of each feedback scheme tested. Up to this point, the hardware description is identical to a standard real-time fiber noise cancellation scheme.

The important difference that our new scheme introduces is in the way the correction signal is processed. The heterodyne signal from the in-loop detector is mixed down in frequency, digitized, and its phase is extracted. This phase contains not only the round-trip fiber noise, but also copies of the correction signal that were applied via the feedback AOM both at the beginning of the outward journey ($2\tau$ ago) and at the end of the return journey (presently). To this digital phase we add two corresponding terms, each proportional to $\kappa$: a negative copy of the feedback signal from $2\tau$ ago and a positive copy of the most recent feedback signal applied. This method requires retaining a rolling record of the feedback signal applied up to $2\tau$ ago, which we accomplish using a field-programmable gate array (FPGA). Weighting the present and $2\tau$-delayed copies of the correction signal in this manner gives the desired temporal shift of $\kappa \tau$ (see Methods for more details). This weighted signal is then conditioned by a digital servo to add frequency-dependent gain and converted back to an analog signal, which is used to drive a voltage-controlled oscillator (VCO) actuating the feedback AOM. Prior knowledge of the transit time $\tau$ is required for this scheme, but this is easily obtainable from interferometric fringes visible on the in-loop spectrum at the local end. Notably, the only additional requirement for this scheme, relative to the standard method, is the digital signal processing; we expect that adapting an existing fiber noise cancellation setup to take advantage of this new method should be relatively low-cost and simple. It is worth noting that a fully analog implementation of this temporal advance scheme is also possible (as outlined in Methods), but requires additional hardware and cannot be as readily adapted to different $\kappa$ values as the digital implementation.


\subsection*{Deployed Fiber Demonstration}

\begin{figure}[!htb]
    \centering
    \makebox[\textwidth][c]{\includegraphics[width=0.8\paperwidth]{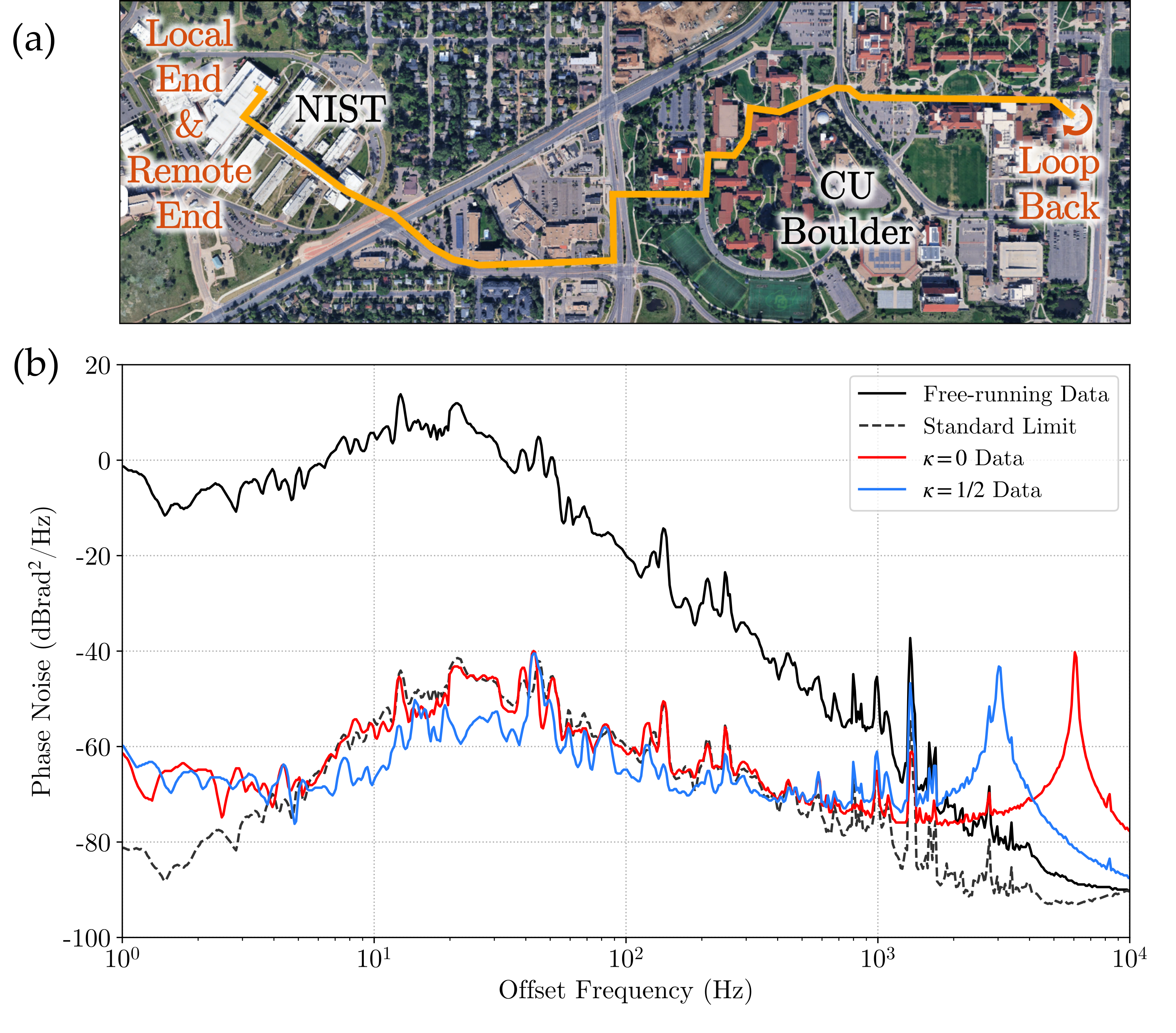}}%
    \caption{\textbf{Performance in deployed fiber.} (a)~Overview map of the section of the BRAN network fiber in Boulder, Colorado used in the experiment. Light is looped back such that both ends of the fiber are co-located at the `Local End'. One-way fiber length is approximately 5.5~km. Map data: Google \textcopyright2024 (b)~Optical phase noise in the standard case ($\kappa=0$), the predicted phase noise limit for the standard case, and phase noise in the optimized case assuming a uniform spatial distribution ($\kappa=1/2$). The noise at offset frequencies below 5 Hz is limited by out-of-loop and non-common fiber paths. The free-running fiber noise is also shown for comparison.}
    \label{BRANfig}
\end{figure}

As a first demonstration of this new feedback scheme, we suppressed the noise in a 5.5~km fiber deployed in an urban environment. This fiber is part of the Boulder Research and Administrative Network (BRAN) in Boulder, Colorado, and stretches from the campus of the National Institute of Standards and Technology (NIST) to the campus of the University of Colorado (CU) Boulder, as shown in Fig.~\ref{BRANfig}a.
To facilitate measurements at both ends of the fiber link path, the local and remote ends were co-located at NIST with a loop back at the CU Boulder side. This was achieved using two separate single mode fibers in a co-packaged fiber bundle. 

Using our digital feedback control scheme, we proceeded to cancel the fiber noise with both the standard $\kappa=0$ delay and with $\kappa=1/2$. Although the exact spatial noise distribution along this deployed fiber is not known, we selected $\kappa=1/2$ under the assumption that the noise is approximately uniform. (In fact, if the noise is nonuniform but mirrored about the midpoint due to the loop back configuration, we would still expect an optimal $\kappa$ of 1/2.) Fig.~\ref{BRANfig}b shows the results of this experiment. In combination with equation~(\ref{equniformlimit}), the free-running noise (noise on the delivered light when no feedback is engaged) defines the limit on the residual fiber noise for a given $\kappa$ value. We show that over a broad range of noise frequencies ($\sim 6-400$~Hz), the new feedback scheme out-performs the standard scheme, which itself is at its theoretical limit. It is worth noting that there is an evident trade-off in bandwidth (discussed more in Methods) as $\kappa$ increases, as seen by the reduction in the frequency of the servo peak from 6~kHz for $\kappa = 0$ to 3~kHz for $\kappa = 1/2$.  However, we find that this is a worthwhile concession, since the frequency range over which we see the most improved performance also contains the highest free-running noise --- meaning the reduction is greatest where it is needed most. This improvement can also be appreciated by examining the integrated phase noise. For $\kappa = 0$, integrating the noise from 10~kHz to 1~Hz (1~kHz to 1~Hz) results in 153.7~mrad (36.2~mrad), whereas for $\kappa = 1/2$, integration over the same frequency range gives 100.6~mrad (25.1~mrad). We also implemented the fully analog scheme over this fiber link, with both phase noise and modified Allan deviation results given in Methods.

\subsection*{Localized Noise Demonstration}

Whereas the previous section demonstrated the utility of this scheme in a deployed fiber, in this section we verify that the optimal temporal advance $\kappa$ coincides with the spatial location of the noise, and we exploit this property to suppress localized noise more than 10 dB beyond the standard limit.
As discussed previously, this new feedback scheme is adaptable to any noise distribution simply by tuning $\kappa$. One class of distributions we can consider is that in which the noise is predominantly localized in one region of the fiber. In this case, it is possible to achieve far better suppression than the standard $\kappa=0$ technique provides. (In principle, perfect cancellation is possible if the noise is entirely in one location. See Methods for more details.) Such a localized noise distribution may arise in an urban link containing an aerial section or in an undersea fiber passing over a geologically active site, for example. 

To model 
a range of localized noise distributions, we built a lab-based fiber testbed to take the place of the `Fiber Link Path' in the Fig.~\ref{digitalsetupfig} diagram. This testbed is an approximately 8.8~km fiber comprised of three long and relatively quiet sections of spooled fiber, plus one short and relatively noisy fiber section that is routed around the lab. In this way, virtually all of the noise is localized within this short section, equivalent to approximately 1\% of the overall fiber length. By changing the order of the component fibers, we can shift where this localized noise is, effectively engineering various noise distributions within our fiber link path. We parameterize the fractional location of the noisy section along the total length in terms of the dimensionless quantity $\alpha$, where $\alpha=0$ ($\alpha=1$) describes a fiber with all of the noise at the local (remote) end.

\begin{figure}[!htb]
    \centering
    \makebox[\textwidth][c]{\includegraphics[width=0.8\paperwidth]{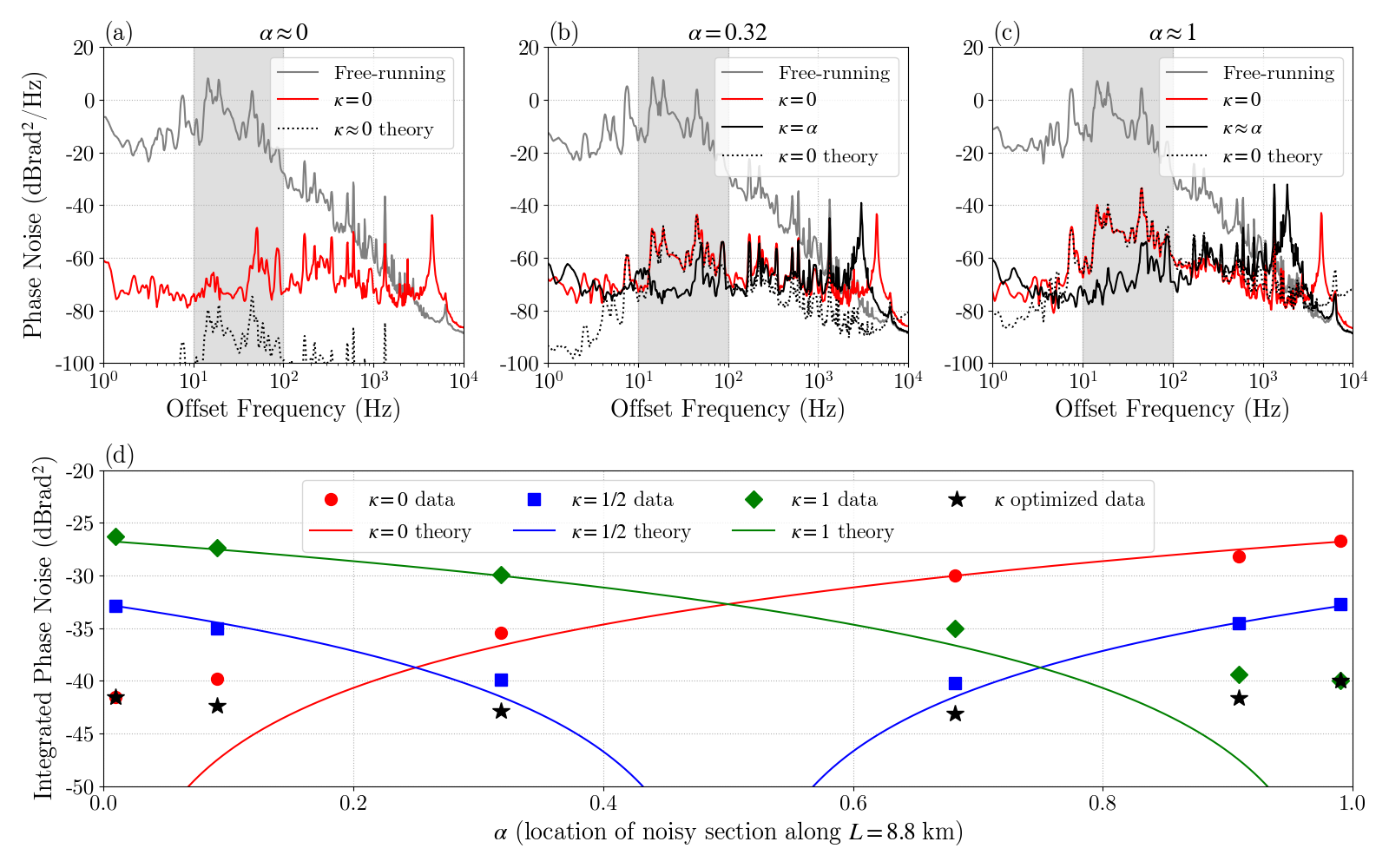}}%
    \caption{\textbf{Optimizing performance for spatially localized noise.} (a-c)~Optical phase noise for fiber configurations where the noise is predominantly localized at the fractional location $\alpha$ along the fiber path. Data is plotted for the standard ($\kappa=0$) case and for the optimized ($\kappa=\alpha$) case. (d)~Comparison of performance for different $\kappa$ values as a function of noise location. Phase noise is integrated from 10-100~Hz (shaded region in (a-c)) for the cases of $\kappa=\{0,~1/2,~1,~\alpha\}$, where $\kappa=\alpha$ gives optimal noise suppression. Since the noisy section of fiber has finite length, $\alpha$ values of 0 and 1 cannot be exactly realized. Based on the fiber lengths used, the first and last configurations are taken to have actual $\alpha$ values of 0.01 and 0.99. Theory curves are derived from equation~(\ref{LN_LF}).}
    \label{localizedfig}
\end{figure}

With this testbed system, we engineered six different fiber noise distributions with the noise localized at different points along the 8.8~km fiber length. Optimal suppression of a localized noise distribution is achieved when $\kappa=\alpha$. Therefore, for each of these distributions, we performed noise cancellation using $\kappa=\{0,~1/2,~1,~\alpha\}$ and measured the out-of-loop phase noise of the light delivered to the remote end in each case. Figs.~\ref{localizedfig}a-c present a selection of these phase noise results focusing on the relative performance of $\kappa=0$ and $\kappa=\alpha$. 
When all of the noise is at the local end of the fiber, $\kappa=0$ gives the optimal performance, since $\alpha$ is approximately zero; however, as the noise is moved further along the fiber (increasing $\alpha$), the performance of the standard $\kappa=0$ case degrades while the $\kappa=\alpha$ case continues to provide optimal noise suppression. Note that the finite length of the noisy section of fiber has led us to approximate the $\alpha$ values of the first and last fiber configurations as 0.01 and 0.99, respectively.

To consolidate these results and the results from the other combinations of $\kappa$ and $\alpha$ we explored, we compare the integrated phase noise over 10-100~Hz for each case, shown in Fig.~\ref{localizedfig}d. This integration range was chosen because it is the peak of the free-running noise PSD (making suppression here most critical) and it is also where the fundamental limits are most consistently the current limitation on performance (allowing for meaningful comparison with theoretical calculations). The theory curves in Fig.~\ref{localizedfig}d are based on equation~(\ref{LN_LF}) in Methods. For every localized noise distribution tested, $\kappa=\alpha$ provides the best noise suppression in this frequency range. In fact, for the cases with the noise in the second half of the fiber (i.e. $\alpha~=~\{0.68,~0.91,~1\}$), the optimized feedback enhances noise suppression by approximately 13~dB compared to the standard $\kappa=0$ case, with further suppression likely possible. With such large suppression of the localized noise, the small amount of noise from the other `quiet' fiber sections will ultimately dominate, once the localized noise is suppressed beyond that lower level. The delay $\kappa$ is only ever perfectly optimized for a single spatial location.

\section*{Discussion}

We have developed a new, intuitive way to conceptualize the correlations between round-trip and one-way noise along a path. By adding a temporal shift through the variable $\kappa$, we introduce a new parameter by which to optimally exploit these correlations in real time. The two experiments presented demonstrate the utility and the potential of this ‘center of mass’ framework for optimally suppressing fiber noise for ultrastable laser delivery. The localized noise experiment supports the assertion that the spatial center of mass of noise along the fiber corresponds to the temporal center of mass of the correlation function. Put differently, localized fluctuations are optimally suppressed when $\kappa=\alpha$, which we show can lead to over 10~dB enhancement in noise suppression relative to the standard limit. The deployed fiber experiment leverages this new understanding to realize a significant improvement in noise suppression in a realistic noise environment, achieving approximately 6~dB enhancement in noise cancellation relative to the standard ($\kappa=0$) case. It is interesting to note that since the center of mass of the correlation function must always lie between $-\tau$ and 0, any $\kappa>0$ will provide enhanced suppression for virtually all possible fiber noise distributions. (Only in the case of noise localized entirely at the very beginning of the fiber path would the standard $\kappa=0$ provide optimal suppression.) That is to say, \textit{any} temporal advance ($\kappa>0$) will provide at least some improvement over the standard scheme for nearly all situations. The optimal temporal advance, of course, corresponds to the noise distribution center of mass. Since the signal processing hardware is the only significant change to the traditional fiber noise cancellation architecture, we expect that the (at least) fourfold performance enhancement presented here can be readily achieved with only minor modification to existing systems.

Using digital signal processing, $\kappa$ can be easily and quickly tuned to any value, making this feedback system extremely adaptable. 
To optimize $\kappa$, the location of the center of mass of the noise distribution is required; this can come from prior knowledge of the fiber path (e.g., the known location of an aerial section), or can be measured with, for example, distributed acoustic sensing (DAS)~\cite{lindsey2019illuminating,liu2025urban}. With DAS, real-time data on the fiber noise could allow dynamic optimization of the noise suppression. This could be of particular advantage in stabilizing paths with noise distributions that change based on tides, rush hour, weather, or any number of external variables. Finally, we note that the framework and feedback schemes developed here need not be limited to optical fiber systems. While optical fiber links represent a ubiquitous and important platform for noise cancellation, these ideas can also be directly applied to free-space optical and microwave systems.

\backmatter

\bmhead{Supplementary information}

Not applicable.

\bmhead{Acknowledgments}

We thank Scott Diddams, Andrew Ludlow, and their respective groups for equipment and access to the BRAN fiber network. We thank Roger Brown, Fabrizio Giorgetta, and Krister Shalm for their helpful comments on the manuscript.

\section*{Declarations}

\begin{itemize}
\item Funding: This research was supported by an appointment to the Intelligence Community Postdoctoral Research Fellowship Program at the National Institute of Standards and Technology administered by Oak Ridge Institute for Science and Education (ORISE) through an interagency agreement between the U.S. Department of Energy and the Office of the Director of National Intelligence (ODNI), and by NIST.

\item Conflict of interest/Competing interests: None.
\item Ethics approval and consent to participate: Not applicable.
\item Data availability: Data available upon reasonable request.
\item Materials availability: Not applicable.
\item Code availability: Not applicable.
\end{itemize}

\begin{appendices}

\section*{Methods}

\subsection*{Digital Servo}

In this section, we provide greater detail on the digital servo architecture, based on the interferometer setup shown in Fig.~\ref{digitalsetupfig}. With an acousto-optic modulator (AOM) driven at frequency $\Omega$, the heterodyne beat frequency on the in-loop detector is $2\Omega$. The phase of this beat signal may be expressed as
\be \theta_0(t) = \varphi_\mathrm{RT}(t) - \theta_c(t) - \theta_c(t-2\tau)  + \varphi_\mathrm{L}(t-2\tau) - \varphi_\mathrm{L}(t), \label{Theta0} \ee
where $\varphi_\mathrm{RT}(t)$ is the round-trip fiber noise, $-\theta_c(t)$ is the correction signal applied to the return light, $-\theta_c (t-2\tau)$ is the correction signal applied to the out-going light after a double-pass through the fiber, $\varphi_\mathrm{L}(t)$ is the laser phase noise from the interferometer’s reference arm, and $\varphi_\mathrm{L}(t-2\tau)$ is the laser phase noise after double-pass through the fiber link. Going forward, we make the usual assumption that the laser source has sufficiently low noise that $\varphi_\mathrm{L}(t-2\tau)-\varphi_\mathrm{L}(t)\approx0$. Satisfying this requirement necessitates a laser with phase noise that is more than 10~dB lower than the free-running phase noise of the fiber link. Most cavity-stabilized lasers meet this requirement for offset frequencies within the lock bandwidth of the fiber noise cancellation. 

At the front end of the servo, the heterodyne signal is digitized and the phase $\theta_0(t)$ is extracted. To achieve the proper weighting of the control signal, $\kappa\theta_c(t)$ and $-\kappa\theta_c(t-2\tau)$ are added to $\theta_0(t)$ digitally. Note that $\kappa$ is chosen by the user and $\tau$ must be known. After signal weighting, the phase is
\be \theta_1(t) = \varphi_\mathrm{RT}(t) - (1-\kappa)\theta_c(t) - (1+\kappa)\theta_c(t-2\tau). \label{Theta1} \ee
In the Fourier domain this becomes
\be \tilde \theta_1(\omega) = \tilde \varphi_\mathrm{RT}(\omega) - \tilde \theta_c(\omega) [(1-\kappa)+(1+\kappa)e^{2i\omega\tau}]. \label{Theta1FD} \ee
An implicit assumption here is that the temporal coherence of $\theta_c$ is greater than the fiber round-trip time, such that the correction applied at $t-2\tau$ can be linearly extrapolated from the signal at time $t$. This holds for noise frequencies where ${\omega\tau\ll1}$, in which case it is equivalent to the first-order approximation of ${\theta_c(t-2\tau)\approx \theta_c(t)+(-2\tau)((d\theta_c)/dt)}$. 

This signal is then conditioned by the digital phase-lock loop (PLL) to provide a frequency-dependent gain, and applied to a voltage-controlled oscillator (VCO) to drive the AOM. The VCO's output frequency is proportional to the control signal input. In terms of the phase, the VCO's output is the derivative of $\theta_c(t)$, such that the Fourier domain equation of the PLL is written as
\be -i\omega \tilde \theta_c (\omega) = F\left(\tilde \varphi_\mathrm{RT}(\omega) - \tilde \theta_c(\omega)[(1-\kappa)+(1+\kappa) e^{2i\omega\tau} ]\right), \label{PLL_FD} \ee
where $F$ is the open-loop gain of the PLL. Solving for $\tilde \theta_c(\omega)$ yields
\be \tilde \theta_c(\omega) = \frac{F/(-i\omega)}{1+\frac{F}{-i\omega}[(1-\kappa)+(1+\kappa)e^{2i\omega\tau}]} \tilde \varphi_\mathrm{RT} (\omega). \label{Thetac_FD} \ee
In the high gain limit where $F\gg1$, this reduces to
\be \tilde \theta_c(\omega) = \frac{1}{(1-\kappa)+(1+\kappa)e^{2i\omega\tau}} \tilde \varphi_\mathrm{RT} (\omega), \label{Thetac_FD2} \ee
which is the desired form. The standard feedback configuration with $\kappa = 0$ can be considered a special case of our more general description where $\kappa$ can vary, typically between 0 and 1. 

In our experiment, we do not directly measure $\theta_c (t)$ and scale by $\kappa$. Rather, we use the digital control signal used to drive the VCO. The control signal was scaled by the volts-to-hertz conversion gain of the VCO, then integrated to retrieve the phase. For our implementation, $\kappa$ was entered from the user-interface of the computer-controlled digital PLL. Memory of the correction signal for a time delay of $2\tau$ is required to apply the $\kappa\theta_c(t-2\tau)$ term. In our system, the delay memory, control signal scaling and integration, and digital PLL are all implemented with an FPGA.
Lastly, we note that the optimal correction applied along the outgoing path is sensitive to errors in the estimated round-trip time. This issue is less critical at low frequencies, because an error $\Delta\tau$ would introduce an error of only $\operatorname{exp}(2i\omega\Delta\tau) - 1 \simeq 2i\omega\Delta\tau$ into the subtracted phase.

\subsection*{Generalized Analog Hardware Implementation}

Analog hardware implementations of the optimized servo architecture are possible, though they are much more cumbersome than the digital servo approach. These implementations require a physical separation of the noise-corrected signal from the noise measurement-and-correction AOM used in the standard scheme. This increases the number of AOMs needed at the local end of the fiber. Also, the drive frequencies of these AOMs are determined by the desired value of $\kappa$, necessitating a change in hardware whenever $\kappa$ is appreciably changed. 

\begin{figure}[!htb]
    \renewcommand\figurename{Extended Data Fig.}
    \centering
    \makebox[\textwidth][c]{\includegraphics[width=0.75\paperwidth]{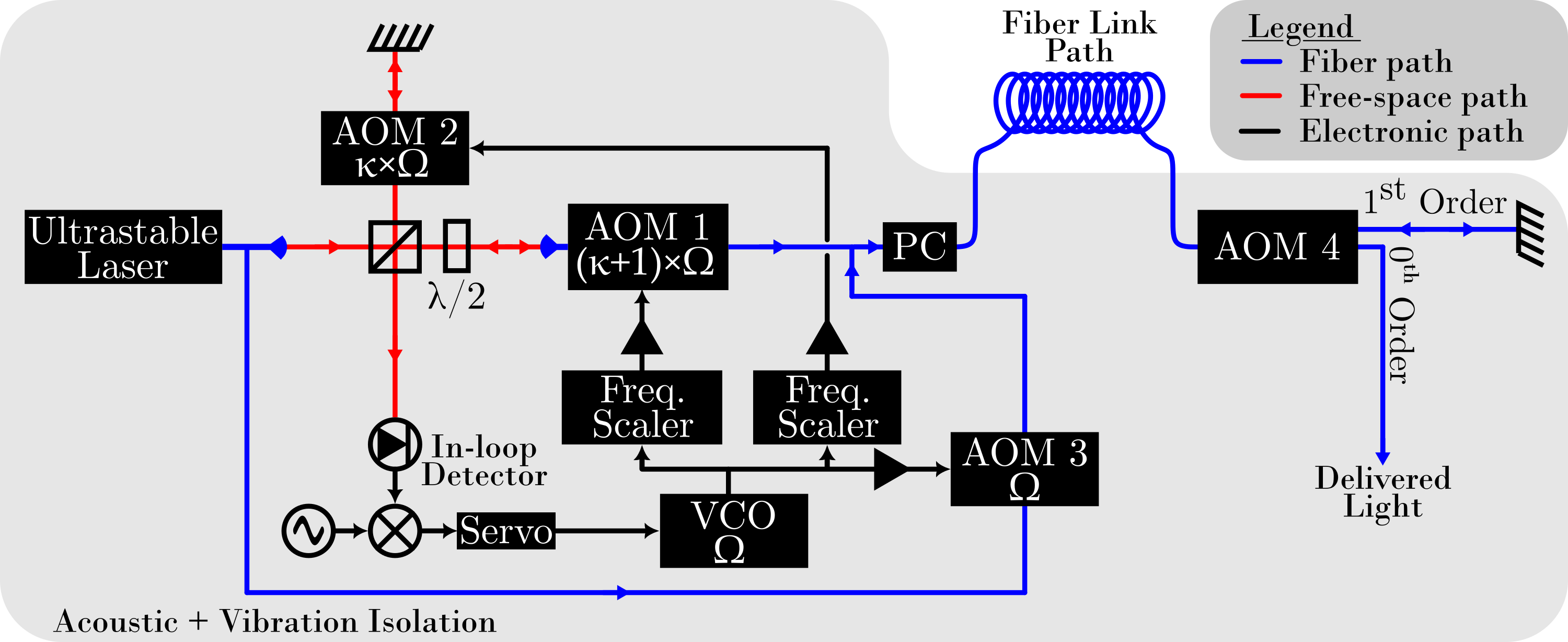}}%
    \caption{\textbf{Setup for analog hardware implementation of the general case.} AOM, acousto-optic modulator; BPF, bandpass filter; VCO, voltage-controlled oscillator; PC, polarization controller.}
    \label{generalizedsetupfig}
\end{figure}

The general scheme for an arbitrary value of $\kappa$ is shown in Extended Data Fig.~\ref{generalizedsetupfig}.
In addition to the double-pass AOM at the fiber input (denoted AOM1), a double-pass AOM is included in the interferometer reference arm (AOM2). Signal weighting is achieved by driving AOM1 and AOM2 with correction signals scaled by $1+\kappa$ and $\kappa$, respectively. This can be achieved, for example, by using a single VCO whose output is split and routed to different frequency scalers. With this arrangement, the phase of signal on the detector from the reference arm becomes $2\kappa\theta_c(t)$, and the signal from the fiber arm is $\varphi_\mathrm{RT} - (1+\kappa) \theta_c(t) - (1+\kappa) \theta_c(t-2\tau)$. The resulting phase of the heterodyne beat on the in-loop detector then matches that of equation~(\ref{Theta1}). 

Neither AOM1 nor AOM2 receives the properly scaled signal for noise cancellation. Therefore, a third AOM (AOM3) is required for delivery of the noise-corrected signal. This signal must be coupled into the fiber after AOM1, adding an uncompensated light path. A fourth AOM is included at the remote end simply to distinguish light that made a full round trip, thereby minimizing the influence of stray reflections. When heterodyning against another laser at the remote end, the different frequency shifts imparted by the separate AOMs provide for straightforward selection of the correct signal. In other use cases, multiple cores of a multicore fiber (MCF) may be used as a separate channel for light delivery. The use of MCF relies on the high degree of noise correlation between the core used to measure the noise and the core used for signal delivery. This has been shown to work for high-fidelity frequency transfer using multiple cores \cite{Hoghooghi:25}. A small benefit of the hardware scheme is that prior knowledge of $\tau$ is not required.

\begin{figure}[!htb]
    \renewcommand\figurename{Extended Data Fig.}
    \centering
    \makebox[\textwidth][c]{\includegraphics[width=0.75\paperwidth]{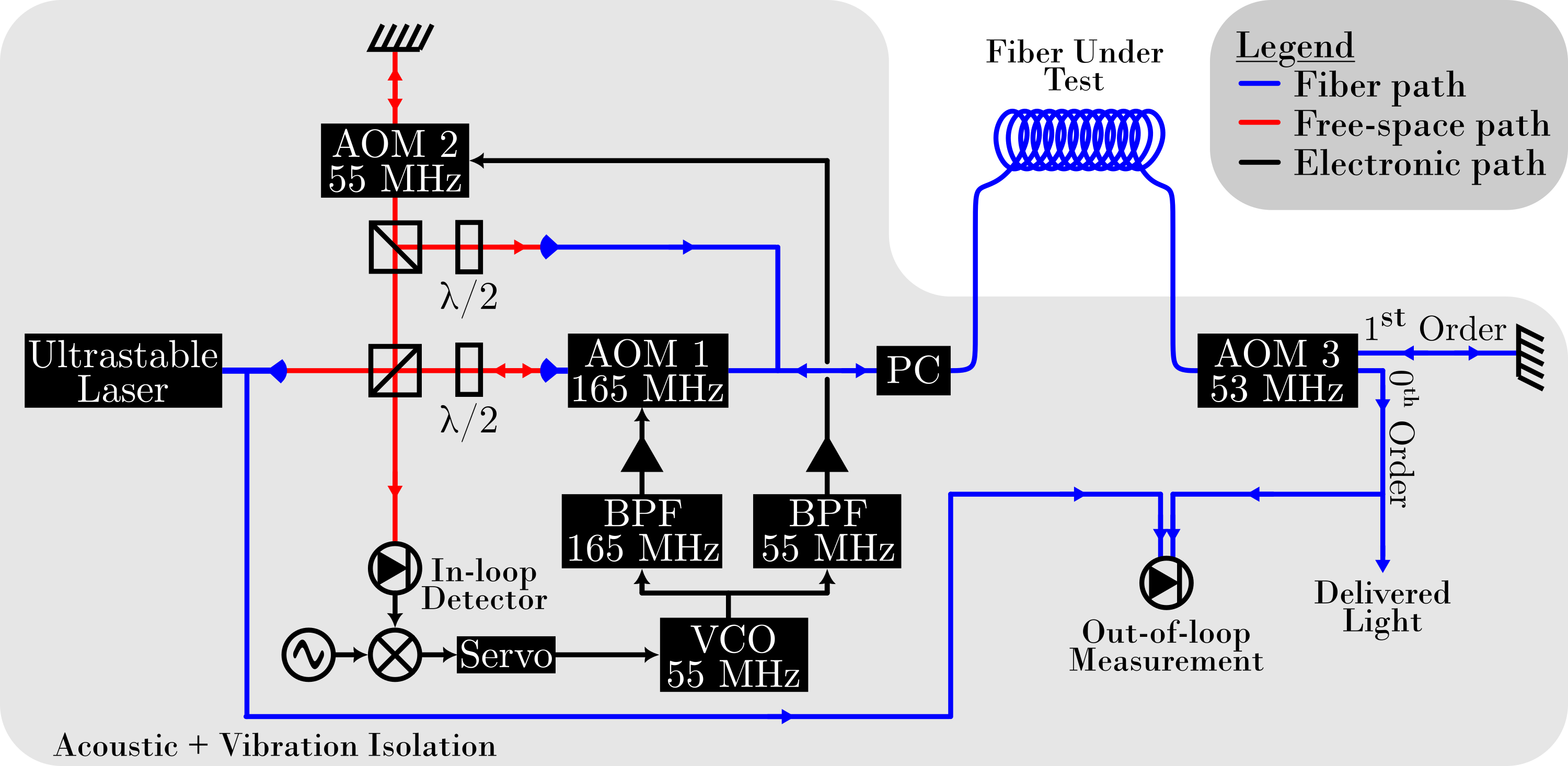}}%
    \caption{\textbf{Setup for hardware implementation of $\kappa=1/2$ case.} AOM, acousto-optic modulator; BPF, bandpass filter; VCO, voltage-controlled oscillator; PC, polarization controller. The two feedback AOM signals are obtained from one VCO by driving the 55~MHz VCO at high power and filtering out the fundamental tone and third harmonic.}
    \label{analogsetupfighalf}
\end{figure}

\begin{figure}[!htb]
    \renewcommand\figurename{Extended Data Fig.}
    \centering
    \makebox[\textwidth][c]{\includegraphics[width=0.8\paperwidth]{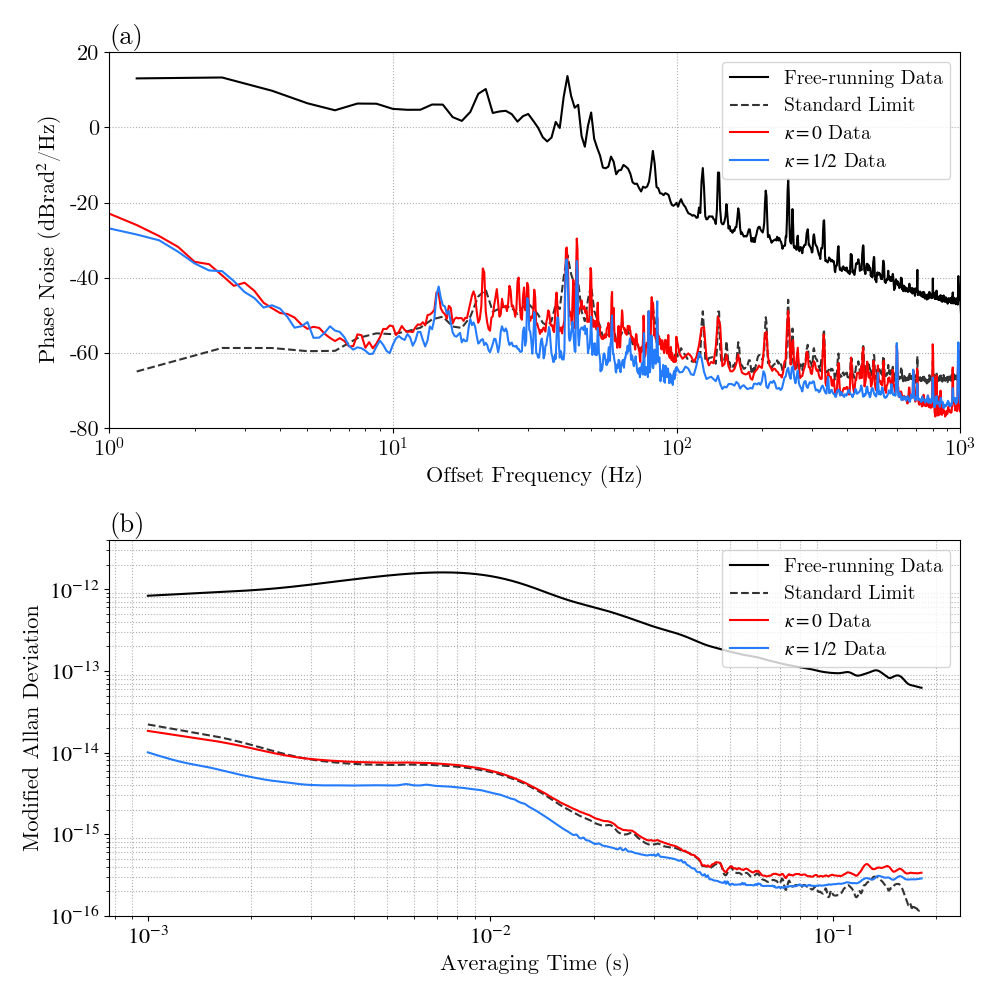}}
    \caption{\textbf{Performance of analog implementation in deployed fiber.} (a)~Optical phase noise in the standard case ($\kappa=0$) and the optimized case ($\kappa=1/2$) for the BRAN deployed fiber. Noise at offset frequencies below 8 Hz is dominated by out of loop and non-common fiber paths. (b)~Modified Allan deviation derived from the phase noise traces in (a).}
    \label{fig_oldBRANdata}
\end{figure}

With $\kappa = 1/2$, the number of required AOMs can be reduced from three to two, since a double-pass through AOM2 provides the correct unity scaling of $\theta_c(t)$. This allows AOM2 to also serve as the delivery AOM. This experimental setup is shown in Extended Data Fig.~\ref{analogsetupfighalf}.
We have successfully realized this analog hardware implementation of $\kappa = 1/2$ to stabilize the BRAN deployed fiber path discussed in the main text. Phase noise and modified Allan deviation results from this experiment are shown in Extended Data Fig.~\ref{fig_oldBRANdata}.
For this demonstration, the reference-arm AOM operated at a frequency of 55~MHz, whereas the in-line AOM operated at 165~MHz, leading to the delivered light having a frequency shift of 110~MHz from double-passing the reference arm AOM. A third AOM was again used in this demonstration at the remote end to distinguish light that made a full round trip.

\subsection*{Lock Bandwidth}

As shown in Figs.~\ref{BRANfig}-\ref{localizedfig}, the locking bandwidth decreases with increasing $\kappa$. In this section, we show how this behavior is expected from the dynamics of the feedback loop. Using standard phase-lock loop analysis~\cite{oppenheim1999discrete}, the in-loop phase may be represented in the frequency domain as 
\be \tilde \theta_1(\omega) = \frac{1}{1+G(\omega)}\tilde \varphi_\mathrm{RT}(\omega), \label{LB1} \ee
\noindent where $G(\omega)$ is given by
\be G(\omega) = \frac{F}{-i\omega}[(1-\kappa)+(1+\kappa)e^{2i\omega\tau}]. \label{Gdef} \ee

The feedback loop will oscillate when $1+G(\omega)\to0$. This can happen when the phase of $G(\omega)$ is equal to $\pi$. Assuming minimal phase delay in $F$, finding the loop oscillation frequency reduces to finding the frequency where 
\be \arg\{[(1-\kappa)+(1+\kappa)e^{2i\omega\tau}]\} = \pi/2 .\label{Garg} \ee

\noindent Defining $\psi =\arg\{[(1-\kappa)+(1+\kappa)e^{2i\omega\tau}]\}$, the equation that gives $\psi$ may be expressed as

\be \tan\psi = \frac{(1+\kappa)\sin(2\omega\tau)}{(1-\kappa)+(1+\kappa)\cos(2\omega\tau)}. \label{psidef} \ee

\begin{figure}[!htb]
    \renewcommand\figurename{Extended Data Fig.}
    \centering
    \makebox[\textwidth][c]{\includegraphics[width=0.8\paperwidth]{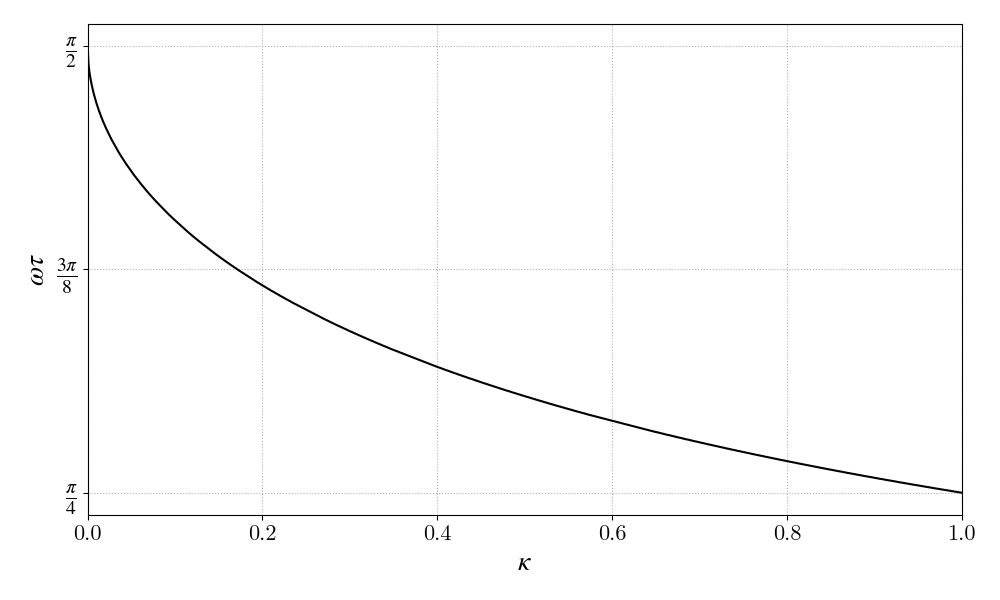}}
    \caption{\textbf{Feedback bandwidth as a function of $\kappa$.} The normalized bandwidth $\omega\tau$ is defined as where the feedback phase is delayed by $\pi$ radians, and is derived by solving equation~(\ref{psidef}).}
    \label{fig_bandwidth}
\end{figure}

When $\kappa = 0$ (standard case), this reduces to $\psi = \omega\tau$.
The frequency for which this phase is equal to $\pi/2$ is therefore $f= 1/(4\tau)$, as expected \cite{Williams:08}. However, when $\kappa = 1$, the phase becomes
$\psi=2\omega\tau$, and the oscillation frequency is reduced to $f = 1/(8\tau)$. The oscillation frequency for intermediate values of $\kappa$ will fall between these extremes. Numerical solutions of equation~(\ref{psidef}) are shown in Extended Data Fig.~\ref{fig_bandwidth}. 

\subsection*{Optimized Noise Suppression: Uniform Distribution}

In this section, we show $\kappa = 1/2$ is the optimal value for noise that is uniformly distributed along the fiber link. A similar conclusion is reached in Ref.~\cite{calosso2015doppler} where it was shown that a temporal advance of the roundtrip noise can be used in post-processing to reduce the fiber noise.

At the remote end of the noise-canceled fiber link, the phase of the signal may be represented as
\be \varphi_\mathrm{remote}(t) = \varphi_\mathrm{fiber}(t)-\theta_c(t-\tau)+\varphi_\mathrm{L}(t-\tau) \label{TDnoiserem}, \ee
where $\varphi_\mathrm{fiber}(t)$ is the one-way noise added by the fiber. 
%
%
%
%
To focus on the link noise, we drop the $\varphi_\mathrm{L}$ term in the analysis below. With this term removed, the Fourier transform of equation~(\ref{TDnoiserem}) is
\be \tilde \varphi_\mathrm{remote}(\omega) = \tilde \varphi_\mathrm{fiber}(\omega)-e^{i\omega\tau}\tilde \theta_c(\omega). \ee
Using the generalized control signal in the high gain limit of equation~(\ref{Thetac_FD2}), the noise at the remote end becomes
\be \tilde \varphi_\mathrm{remote}(\omega) = \tilde \varphi_\mathrm{fiber}(\omega)-\frac{e^{i\omega\tau}}{(1-\kappa)+(1+\kappa)e^{2i\omega\tau}}\tilde \varphi_\mathrm{RT}(\omega). \ee
This simplifies to
\be \tilde \varphi_\mathrm{remote}(\omega) = \tilde \varphi_\mathrm{fiber}(\omega)-\frac{1}{2}\frac{1}{\cos(\omega\tau)+i\kappa\sin(\omega\tau)}\tilde \varphi_\mathrm{RT}(\omega). \label{FDNCF} \ee
%
Further analysis requires expressions for $\tilde \varphi_\mathrm{fiber}(\omega)$ and $\tilde \varphi_\mathrm{RT}(\omega)$. We follow the typical representation of the one-way noise~\cite{Williams:08}, where phase fluctuations imparted at position $z$ (measured from the local end) are represented as $\delta\varphi[z,t-(\tau-z/c_n)]$, where $c_n$ is the speed of light in the fiber. These fluctuations are integrated over $L$, the length of the fiber, yielding
\be \varphi_\mathrm{fiber}(t) = \int_0^L \delta \varphi_z[z, t-(\tau -z/c_n)] \operatorname{dz}. \label{TDf} \ee
Taking the Fourier transform yields
\be \tilde \varphi_\mathrm{fiber}(\omega)=e^{i\omega\tau} \int_0^L e^{-i\omega z/c_n} \delta \tilde \varphi(z,\omega) \operatorname{dz}. \label{FDf} \ee

The round-trip noise consists of phase perturbations on the return light as well as the out-going light after a double-pass, and is likewise integrated over the length of the fiber, yielding
\be \varphi_\mathrm{RT}(t) = \int_0^L \{ \delta \varphi_z(z, t- z/c_n) + \delta \varphi_z[z, t-(2 \tau -z/c_n)] \} \operatorname{dz}, \label{TDRT} \ee
with Fourier transform
\be \tilde \varphi_\mathrm{RT}(\omega) = 2e^{i\omega\tau} \int_0^L \cos[\omega(\tau-z/c_n)] \delta \tilde \varphi (z,\omega) \operatorname{dz}. \label{FDRT}\ee
Inserting equations~(\ref{FDf}) and (\ref{FDRT}) into equation~(\ref{FDNCF}) yields
\be \tilde \varphi_\mathrm{remote}(\omega) = e^{i\omega\tau} \int_0^L \Phi(\omega, z) \delta \tilde \varphi (\omega, z) \operatorname{dz}, \label{FDNCF2} \ee
where
\be \Phi(\omega, z) = e^{-i\omega z/c_n}-\frac{\cos[\omega(\tau-z/c_n)]}{\cos(\omega\tau)+i\kappa\sin(\omega\tau)}. \ee
%
The power spectral density (PSD) of the noise at the remote end is given by ${S_\mathrm{remote}=\langle |\tilde \varphi_\mathrm{remote}(\omega)|^2 \rangle}$. This is expressed as
\be 
\langle |\tilde \varphi_\mathrm{remote}(\omega)|^2 \rangle = \int_0^L \operatorname{dz}' \int_0^L \operatorname{dz} \Phi^*(\omega,z')\Phi(\omega,z) \langle \delta \tilde \varphi^*(\omega, z') \delta \tilde \varphi(\omega,z) \rangle. \label{NCFPSD1} \ee
Note that * here denotes complex conjugation. We now specify the spatial correlation of the fiber noise as
\be 
\langle \delta \tilde \varphi^*(\omega, z') \delta \tilde \varphi(\omega,z)\rangle = C(|z-z'|)\langle|\delta \varphi_z(\omega)|^2\rangle, \ee
where $C(|z-z'|)$ is a dimensionless, smooth function that goes to zero sufficiently fast as $z-z'\to\infty$, and $C(0) = 1$. We further define the characteristic length $L_\mathrm{corr}$ of the spatial correlations as 
\be \int_{-\infty}^\infty C(|z-z'|) \operatorname{dz}' = L_\mathrm{corr}. \ee

If we now assume $L_\mathrm{corr}\ll L$, the function $L_\mathrm{corr}^{-1}C(|z-z'|)$ can be replaced by a Dirac delta function,
\be C(|z-z'|) = L_\mathrm{corr}\delta(z-z'). \ee
Equation (\ref{NCFPSD1}) then simplifies to
\be \langle |\tilde \varphi_\mathrm{remote}(\omega)|^2 \rangle = L_\mathrm{corr} \int_0^L |\Phi(\omega,z)|^2 \langle|\delta \varphi_z(\omega)|^2\rangle \operatorname{dz}. \ee
With spatially uniform noise, this becomes
\be \langle |\tilde \varphi_\mathrm{remote}(\omega)|^2 \rangle = \frac{1}{L}S_\mathrm{fiber}(\omega) \int_0^L |\Phi(\omega,z)|^2 \operatorname{dz}, \label{NCF3}  \ee
where we have used the fact that the PSD of the free-running one-way fiber noise is given by
\be S_\mathrm{fiber}(\omega) =L\cdot L_\mathrm{corr} \langle|\delta \varphi_z(\omega)|^2\rangle. \ee
Straightforward integration of equation~(\ref{NCF3}) yields
\be \langle |\tilde \varphi_\mathrm{remote}(\omega)|^2 \rangle = \frac{1+2\kappa(\kappa-1)\sin^2(\omega\tau)-\operatorname{sinc}(2\omega\tau)}{2[\kappa^2+(1-\kappa^2)\cos^2(\omega\tau)]}S_\mathrm{fiber}(\omega). \ee
%

\noindent Expanding to lowest order in $\omega\tau$, we obtain
\be S_\mathrm{remote}= \langle |\tilde \varphi_\mathrm{remote}(\omega)|^2 \rangle = \left[\kappa^2-\kappa+\frac 1 3 \right](\omega\tau)^2 S_\mathrm{fiber}(\omega), \ee
which is presented as equation~(\ref{equniformlimit}) in the main text. The noise is minimized for $\kappa=1/2$,

\begin{equation}
   S_\mathrm{remote}(\omega)=\frac{1}{12}(\omega \tau)^2S_\mathrm{fiber}(\omega), \quad \kappa=1/2,
   \label{eq12limit}
\end{equation}

\noindent while $\kappa = 0$ yields the standard result,

\begin{equation}
   S_\mathrm{remote}(\omega)=\frac{1}{3}(\omega \tau)^2S_\mathrm{fiber}(\omega), \quad \kappa=0.
   \label{eq3limit}
\end{equation}

\noindent Equation~(\ref{eq12limit}) has previously been derived for two-way transfer and systems utilizing post-processing~\cite{calosso2015doppler,Stefani:15}, both of which require ultrastable lasers at each end of the fiber. Here we show that the limit imposed by equation~(\ref{eq12limit}) is achievable with one-way frequency transfer in real time.

%

\subsection*{Effective Temporal Shift}

We now examine the effective temporal shift of the round-trip noise at the remote end of the fiber when feedback is applied. For $\omega\tau\ll 1$, equation~(\ref{FDNCF}) can be expanded as 
\be \tilde \varphi_\mathrm{remote}(\omega) \approx \tilde \varphi_\mathrm{fiber}(\omega)-\frac{1}{2}\frac{1}{(1+i\kappa\omega\tau)}\tilde \varphi_\mathrm{RT}(\omega)\approx \tilde \varphi_\mathrm{fiber}(\omega)-\frac 12 e^{-i\kappa\omega\tau}\tilde \varphi_\mathrm{RT}(\omega) \ee
In the time domain this is
\be \varphi_\mathrm{remote}(t)\approx \varphi_\mathrm{fiber}(t)-\frac 12 \varphi_\mathrm{RT}(t+\kappa\tau) \ee
Note that since the correction signal $\theta_c(t)$ originates at the local end it is delayed by $\tau$ as it propagates to the remote end of the fiber. The temporal advance of the round-trip noise at the local end is therefore $(\kappa+1)\tau$ (this can be derived directly from equation~(\ref{Thetac_FD2}) as well). Thus, even for the standard $\kappa = 0$ architecture, the local-end correction signal exhibits an effective temporal advance.

\subsection*{Noise Center of Mass}

The main text provides a pictorial illustration of the noise correlation function between the opposite ends of the fiber. In this section, we provide a more rigorous derivation. Using a $1/\sqrt{T}$ normalization for a finite-time Fourier transform over a large time window $T$, the cross-correlation $\langle \varphi_\mathrm{RT}(t - t')  \varphi_\mathrm{fiber}(t) \rangle$ between the phase noise of the out-going and the round-trip fields is the inverse Fourier transform of $\langle \tilde \varphi_\mathrm{RT}^*(\omega) \tilde \varphi_\mathrm{fiber}(\omega) \rangle$. Using equations~(\ref{FDf}) and (\ref{FDRT}), this becomes
\be \langle \tilde \varphi_\mathrm{RT}^*(\omega) \tilde \varphi_\mathrm{fiber}(\omega) \rangle = 2 L_\mathrm{corr} \langle |\delta \tilde \varphi_z (\omega)|^2 \rangle \int_0^L e^{-i \omega z/c_n} \cos[\omega(\tau - z/c_n)] \mathrm{dz}.  \ee
After integration, the cross-spectrum becomes
\be \langle \tilde \varphi_\mathrm{RT}^*(\omega) \tilde \varphi_\mathrm{fiber}(\omega) \rangle = L \left[e^{-i \omega \tau} + \operatorname{sinc}(\omega \tau) \right] L_\mathrm{corr} \langle |\delta \tilde \varphi_z (\omega)|^2 \rangle .  \ee
Inverse Fourier transform of the above produces the convolution between
\be K(t') = L \left[\delta(t'+\tau) + \frac 1 {2 \tau} \operatorname{rect}\left( \frac {t'} {2 \tau} \right)\right] \ee
and the inverse Fourier transform of $L_\mathrm{corr} \langle |\delta \tilde \varphi_z (\omega)|^2 \rangle$, which is the temporal correlation function of the phase perturbations. Here, $\operatorname{rect}(\cdot)$ is a rectangular function of unit amplitude and unit width. Assuming that the fiber perturbations are much faster than $\tau$, their temporal correlation function can be approximated by a delta function. As a result, we obtain
\be \langle \varphi_\mathrm{RT}(t - t') \varphi_\mathrm{fiber}(t) \rangle = L \left[\delta(t'+\tau) + \frac 1 {2 \tau} \operatorname{rect}\left( \frac {t'} {2 \tau} \right)\right] L_\mathrm{corr} \langle |\delta \tilde \varphi_z (0)|^2 \rangle. \ee

The function $K(t')$, sketched in Fig.~\ref{fig_intuitive}c of the main text, represents the effect of a delay $t'$ on the temporal overlap between phase noise $\varphi_\mathrm{fiber}(t)$ and $\varphi_\mathrm{RT}(t-t')$. The two terms of the correlation function have the same area, with the first term centered on $t'=-\tau$ and the second on $t'=0$. The total temporal correlation center of mass is therefore centered on $t'=-\tau/2$. The minus sign indicates a delay of $\varphi_\mathrm{RT}(t)$ relative to $\varphi_\mathrm{fiber}(t)$. The optimal correction signal therefore requires a temporal advance of $\tau/2$. In a system where the noise is band limited, the inverse Fourier transform of $L_\mathrm{corr} \langle |\delta \tilde \varphi_z (\omega)|^2 \rangle$ is not a delta function. However, the noise center of mass does not change.

\subsection*{Optimized Noise Suppression: Localized Fluctuations}
In this section, we derive the expected noise suppression for the situation where the phase noise fluctuations are dominated by a specific section of the fiber. Examples include fibers with aerial sections, near the shore of undersea cables, or near geologically active sites. 

For noise located at $z_0$ measured from the local end, we first define $\alpha$ as the fractional length to the noise location, or $\alpha = z_0/L$. We limit the spatial extent of the fiber noise over a length $L_0$ by defining $\delta\tilde\varphi(z,\omega)$ as 
\be \delta\tilde\varphi(z,\omega)=\delta\tilde\varphi_z(\omega)\operatorname{rect}[(z-z_0)/L_0] \ee
If we then assume $\Phi(\omega, z)$ is constant over $L_0$, the PSD of the noise at the remote end becomes
\be S_\mathrm{remote}(\omega)=\left|e^{-i\omega\tau\alpha}- \frac{\cos[\omega\tau(1-\alpha)]}{\cos(\omega\tau)+i\kappa\sin(\omega\tau)}\right|^2 S_\mathrm{fiber}(\omega) \ee
Note that the noise is fully suppressed for $\kappa=\alpha=\{0,1\}$. (Recall that this applies in the high gain limit in steady state.) Expansion to second order gives
\begin{multline}
    S_\mathrm{remote}(\omega) \approx (\kappa - \alpha)^2 (\omega \tau)^2 + (\omega \tau)^4 \big[ (\kappa^2 - \alpha)^2 \\ -\frac 1 3  (\alpha-\kappa) (2 \kappa + \alpha^2 - 3 \kappa(2 \kappa^2 + \alpha-2))\big] S_\mathrm{fiber}(\omega).
    \label{LN_LF}
\end{multline}
%
The lowest noise is achieved when $\kappa=\alpha$, where the PSD is reduced to 
\be  S_\mathrm{remote}(\omega)\approx \alpha^2 (\alpha - 1)^2 (\omega \tau)^4 S_\mathrm{fiber}(\omega). \ee
The maximum residual noise is when $\alpha$ is in the center of the fiber link, where the noise is $(1/16)(\omega\tau)^4$. In practice, whenever $\kappa = \alpha$, the localized suppression is so great that the noise from other sections of the fiber will usually dominate at low frequencies. 

With digital signal processing, it is possible to further improve this limit for localized noise by adding another term to the heterodyne phase with delay of $2 \alpha \tau$. In this case, it is theoretically possible to achieve perfect cancellation of noise for any value of $\alpha$, so long as the fiber noise is perfectly localized. The phase of the weighted signal of equation~(\ref{Theta1}) then becomes


\be \theta_1(t) = \varphi_\mathrm{RT}(t) - (1-\kappa_1)\theta_c(t) - (1+\kappa_2)\theta_c(t-2\tau) - \theta_c(t-2\alpha\tau). \label{Theta1LN}
\ee

\noindent We now set $\kappa_1 = 1$ and $\kappa_2 = 0$, yielding

\be \theta_1(t) = \varphi_\mathrm{RT}(t) - \theta_c(t-2\tau) - \theta_c(t-2\alpha\tau). \label{Theta1LN2}
\ee

\noindent In the high gain limit, the frequency domain representation of the correction signal is

\be \tilde \theta_c(\omega) = \frac{1}{e^{2i\omega\tau}+e^{2i\omega\alpha\tau}}\tilde \varphi_\mathrm{RT}(\omega).
\ee

\noindent The noise-corrected PSD at the remote end of the fiber is therefore
\be S_\mathrm{remote}(\omega) = \left| e^{-i\omega\alpha\tau}-\frac{2\cos[\omega\tau(1-\alpha)]}{e^{i\omega\tau}+e^{i\omega\tau(2\alpha-1)}} \right|^2 S_\mathrm{fiber}(\omega),
\ee

\noindent which is zero for all values of $\alpha$.

It is interesting to note that the average of the term proportional to $(\omega\tau)^2$ in equation~(\ref{LN_LF}) with a uniform distribution of the position of the noise localized at $\alpha$ is 
\bea \int_0^1 \frac{S_\mathrm{remote}(\omega)}{S_\mathrm{fiber}(\omega)} \operatorname{d \alpha} &\simeq& (\omega \tau)^2 \int_0^1 (\kappa - \alpha)^2 \operatorname{d\alpha} \nonumber \\ &=& \left[\kappa^2 - \kappa + \frac{1}{3}\right] (\omega \tau)^2, \eea
producing once again that the optimum value of $\kappa$ is one half. The impossibility of suppressing the term proportional to $(\omega \tau)^2$ for uncorrelated fluctuations can be interpreted as a consequence of the fact that the value of $\kappa$ that produces optimal suppression depends on the position of the perturbations along the fiber. For spatially distributed noise, it is impossible to pick one value of $\kappa$ that perfectly suppresses noise at multiple locations simultaneously. Therefore, the best compromise comes from choosing a $\kappa$ that corresponds to the noise distribution center of mass.

\subsection*{Spatially Correlated Noise}
The phase noise may be correlated along the fiber path, as is expected in the use of spooled fiber. Here we show that, when the fluctuations are fully correlated in space, the use of the proposed technique with $\kappa = 1/2$ is much more effective than the standard $\kappa=0$ architecture. In this case, there is no spatial dependence to the phase fluctuations $\delta\tilde\varphi(z,\omega)=\delta \tilde\varphi_z(\omega)$, hence equation~(\ref{FDNCF2}) can be explicitly integrated, yielding
\be \tilde \varphi_\mathrm{remote}(\omega)= \tilde \varphi_\mathrm{fiber}(\omega) \bigg[e^{-i\omega\tau/2} \operatorname{sinc}(\omega \tau/2) - \frac{2 \operatorname{sinc}(\omega \tau)}{(1-\kappa) e^{-i\omega\tau} + (1+\kappa)e^{i\omega\tau}}\bigg],
\ee
where we have used $L\delta\tilde\varphi_z(\omega) = \tilde\varphi_\mathrm{fiber}(\omega)$. This holds because of the spatial coherence of the fluctuations. Taking the magnitude squared and expanding for small $\omega\tau$ we obtain

\bea S_\mathrm{remote}(\omega)  &=&  \frac 1 4 \big[ (\omega \tau)^2 (2 \kappa - 1)^2 \nonumber \\ && \hspace{-1cm} + \frac 1 {6} (\omega \tau)^4 (- 24\kappa^4 + 24\kappa^3 + 8\kappa^2 - 14\kappa + 5) \big]S_\mathrm{fiber}(\omega). \label{uncorr}\eea

For $\kappa = 0$, the result is nearly the same as it is for uniformly distributed, uncorrelated noise, with only the $1/3$ prefactor replaced with $1/4$. For $\kappa = 1/2$, however, the $(\omega\tau)^2$ term falls away, and we obtain the qualitatively different result of
\be
S_\mathrm{remote}(\omega) = \frac {1}{16} (\omega\tau)^4 S_\mathrm{fiber}(\omega).
\ee
This can lead to drastically lower noise. For example, for $\tau =$ 50 $\mu$s (10 km of fiber), $\kappa = 1/2$ results in an extra 36 dB of suppression of the fiber noise at 100 Hz offset compared to $\kappa = 0$.

\subsection*{Optimality of the Scheme}
In equation~(\ref{eqnewscheme}), the control signal is written as the round-trip noise scaled by $1/2$ and temporally shifted by $\kappa\tau$. The most general form of the control signal is a linear transformation of the round-trip noise, such that the noise at the remote end of the fiber may be written as
\be  \tilde \varphi_\mathrm{remote}(t) = \varphi_\mathrm{fiber}(t)-h(t)*\varphi_\mathrm{RT}(t), \ee
where $h(t)$ is the impulse response of the control system after propagating to the remote end of the fiber, and $*$ denotes convolution. The most general form of the control kernel $\Phi(\omega,z)$ is therefore
\be \Phi(\omega,z) = e^{-i\omega z/c_n} - H(\omega)\cos[\omega(\tau-z/c_n)], \label{optPhi} \ee
where $H(\omega)$ is the Fourier transform of $h(t)$. To ensure full noise cancellation at $\omega = 0$, the condition $\Phi(0,z) = 0$ must be satisfied. In this case, a useful form of $H(\omega)$ is
\be H(\omega) = 1 + (\omega\tau)\chi(\omega\tau), \label{Hdef} \ee
where $\chi(\cdot)$ is an arbitrary, generally complex, function of its argument. In this section, we show our scheme produces the value of this function that results in the lowest noise.
We restrict our discussion to the case of uncorrelated, uniformly distributed noise. For localized noise, the noise can in principle be fully suppressed. For spatially correlated noise, at small $\omega\tau$ the $(\omega\tau)^4$ dependence renders this noise negligible. Performing the integral in equation~(\ref{FDNCF2}) with the use of equations~(\ref{optPhi}) and (\ref{Hdef}) and expanding to second order gives
\be \frac{\langle |\tilde \varphi_\mathrm{remote}(\omega)|^2 \rangle}{\langle |\tilde \varphi_\mathrm{fiber}(\omega)|^2 \rangle} = \left[\frac 1 3 -\operatorname{imag}\chi(0) +  |\chi(0)|^2 \right](\omega \tau)^2 + O[(\omega \tau)^4]. \label{eqoptimal} \ee
Here we have used the fact that only $\chi(0)$ affects the second-order term. A nonzero real part of $\chi(0)$ can only increase the noise, so that we may assume  $\chi(0)$ as purely imaginary: $\chi(0)= i \kappa$. With this choice, equation~(\ref{eqoptimal}) becomes
\be \frac{\langle |\tilde \varphi_\mathrm{remote}(\omega)|^2 \rangle}{\langle |\tilde \varphi_\mathrm{fiber}(\omega)|^2 \rangle} = \left(\frac 1 3 - \kappa +  \kappa^2 \right)(\omega \tau)^2 + O[(\omega \tau)^4]. \ee
Thus, the minimum noise is achieved for $\kappa = 1/2$ and is equal to $(\omega \tau)^2/12$. It is impossible to obtain, for spatially uncorrelated perturbations, a faster decay of the noise spectrum at low frequency for any choice of $\chi(\omega\tau)$.

\end{appendices}

\bibliography{sn-bibliography.bib}

@article{Williams:08,
author = {P. A. Williams and W. C. Swann and N. R. Newbury},
journal = {J. Opt. Soc. Am. B},
number = {8},
pages = {1284--1293},
publisher = {Optica Publishing Group},
title = {High-stability transfer of an optical frequency over long fiber-optic links},
volume = {25},
month = {Aug},
year = {2008},
url = {https://opg.optica.org/josab/abstract.cfm?URI=josab-25-8-1284},
doi = {10.1364/JOSAB.25.001284},
}

@article{Ma:94,
author = {Long-Sheng Ma and Peter Jungner and Jun Ye and John L. Hall},
journal = {Opt. Lett.},
number = {21},
pages = {1777--1779},
publisher = {Optica Publishing Group},
title = {Delivering the same optical frequency at two places: accurate cancellation of phase noise introduced by an optical fiber or other time-varying path},
volume = {19},
month = {Nov},
year = {1994},
url = {https://opg.optica.org/ol/abstract.cfm?URI=ol-19-21-1777},
doi = {10.1364/OL.19.001777},
}

@article{calosso2015doppler,
  title={Doppler-stabilized fiber link with 6~d{B} noise improvement below the classical limit},
  author={Calosso, CE and Bertacco, EK and Calonico, D and Clivati, C and Costanzo, Giovanni Antonio and Frittelli, M and Levi, F and Micalizio, S and Mura, A and Godone, A},
  journal={Optics Letters},
  volume={40},
  number={2},
  pages={131--134},
  year={2015},
  publisher={Optical Society of America}
}

@article{Stefani:15,
author = {Fabio Stefani and Olivier Lopez and Anthony Bercy and Won-Kyu Lee and Christian Chardonnet and Giorgio Santarelli and Paul-Eric Pottie and Anne Amy-Klein},
journal = {J. Opt. Soc. Am. B},
number = {5},
pages = {787--797},
publisher = {Optica Publishing Group},
title = {Tackling the limits of optical fiber links},
volume = {32},
month = {May},
year = {2015},
url = {https://opg.optica.org/josab/abstract.cfm?URI=josab-32-5-787},
doi = {10.1364/JOSAB.32.000787},
}

@article{Hoghooghi:25,
author = {Nazanin Hoghooghi and Mikael Mazur and Nicolas Fontaine and Yifan Liu and Dahyeon Lee and Charles McLemore and Takuma Nakamura and Tetsuya Hayashi and Giammarco Di Sciullo and Divya Shaji and Antonio Mecozzi and Cristian Antonelli and Franklyn Quinlan},
journal = {Optica},
number = {6},
pages = {894--899},
publisher = {Optica Publishing Group},
title = {Ultrastable optical frequency transfer and attosecond timing in deployed multicore fiber},
volume = {12},
month = {Jun},
year = {2025},
url = {https://opg.optica.org/optica/abstract.cfm?URI=optica-12-6-894},
doi = {10.1364/OPTICA.558821},
}

@article{marra2018ultrastable,
  title={Ultrastable laser interferometry for earthquake detection with terrestrial and submarine cables},
  author={Marra, Giuseppe and Clivati, Cecilia and Luckett, Richard and Tampellini, Anna and Kronj{\"a}ger, Jochen and Wright, Louise and Mura, Alberto and Levi, Filippo and Robinson, Stephen and Xuereb, Andr{\'e} and others},
  journal={Science},
  volume={361},
  number={6401},
  pages={486--490},
  year={2018},
  publisher={American Association for the Advancement of Science}
}

@article{marra2022optical,
  title={Optical interferometry--based array of seafloor environmental sensors using a transoceanic submarine cable},
  author={Marra, G and Fairweather, DM and Kamalov, V and Gaynor, P and Cantono, M and Mulholland, S and Baptie, B and Castellanos, JC and Vagenas, G and Gaudron, J-O and others},
  journal={Science},
  volume={376},
  number={6595},
  pages={874--879},
  year={2022},
  publisher={American Association for the Advancement of Science}
}

@article{manley2023searching,
  title={Searching for scalar ultralight dark matter via refractive index changes in fibers},
  author={Manley, J and Stump, R and Petery, R and Singh, S},
  journal={Physical Review D},
  volume={108},
  number={7},
  pages={075008},
  year={2023},
  publisher={APS}
}

@article{roberts2020search,
  title={Search for transient variations of the fine structure constant and dark matter using fiber-linked optical atomic clocks},
  author={Roberts, Benjamin M and Delva, Pacome and Al-Masoudi, Ali and Amy-Klein, Anne and Baerentsen, Christian and Baynham, CFA and Benkler, Erik and Bilicki, Slawomir and Bize, Sebastien and Bowden, William and others},
  journal={New Journal of Physics},
  volume={22},
  number={9},
  pages={093010},
  year={2020},
  publisher={IOP Publishing}
}

@article{lindsey2021fiber,
  title={Fiber-optic seismology},
  author={Lindsey, Nathaniel J and Martin, Eileen R},
  journal={Annual Review of Earth and Planetary Sciences},
  volume={49},
  number={1},
  pages={309--336},
  year={2021},
  publisher={Annual Reviews}
}

@article{mehlstaubler2018atomic,
  title={Atomic clocks for geodesy},
  author={Mehlst{\"a}ubler, Tanja E and Grosche, Gesine and Lisdat, Christian and Schmidt, Piet O and Denker, Heiner},
  journal={Reports on Progress in Physics},
  volume={81},
  number={6},
  pages={064401},
  year={2018},
  publisher={IOP Publishing}
}

@inproceedings{cecilia2015fiber,
  title={A fiber link for the remote comparison of optical clocks and geodesy experiments},
  author={Clivati, Cecilia and Calonico, Davide and Frittelli, Matteo and Mura, Alberto and Levi, Filippo},
  booktitle={2015 Joint Conference of the IEEE International Frequency Control Symposium \& the European Frequency and Time Forum},
  pages={579--582},
  year={2015},
  organization={IEEE}
}

@article{grotti2018geodesy,
  title={Geodesy and metrology with a transportable optical clock},
  author={Grotti, Jacopo and Koller, Silvio and Vogt, Stefan and H{\"a}fner, Sebastian and Sterr, Uwe and Lisdat, Christian and Denker, Heiner and Voigt, Christian and Timmen, Ludger and Rolland, Antoine and others},
  journal={Nature Physics},
  volume={14},
  number={5},
  pages={437--441},
  year={2018},
  publisher={Nature Publishing Group UK London}
}

@article{collaboration2021frequency,
  title={Frequency ratio measurements at 18-digit accuracy using an optical clock network},
  author={{BACON Collaboration} and others},
  journal={Nature},
  volume={591},
  number={7851},
  pages={564--569},
  year={2021}
}

@article{rosenband2008frequency,
  title={Frequency ratio of {A}l$^+$ and {H}g$^+$ single-ion optical clocks; metrology at the 17th decimal place},
  author={Rosenband, Till and Hume, DB and Schmidt, PO and Chou, Chin-Wen and Brusch, Anders and Lorini, Luca and Oskay, WH and Drullinger, Robert E and Fortier, Tara M and Stalnaker, Jason E and others},
  journal={Science},
  volume={319},
  number={5871},
  pages={1808--1812},
  year={2008},
  publisher={American Association for the Advancement of Science}
}

@article{nemitz2016frequency,
  title={Frequency ratio of {Y}b and {S}r clocks with $5\times 10^{-17}$ uncertainty at 150 seconds averaging time},
  author={Nemitz, Nils and Ohkubo, Takuya and Takamoto, Masao and Ushijima, Ichiro and Das, Manoj and Ohmae, Noriaki and Katori, Hidetoshi},
  journal={Nature Photonics},
  volume={10},
  number={4},
  pages={258--261},
  year={2016},
  publisher={Nature Publishing Group}
}

@article{dorscher2021optical,
  title={Optical frequency ratio of a $^{171}${Y}b$^+$ single-ion clock and a $^{87}${S}r lattice clock},
  author={D{\"o}rscher, S{\"o}ren and Huntemann, Nils and Schwarz, Roman and Lange, Richard and Benkler, Erik and Lipphardt, Burghard and Sterr, Uwe and Peik, Ekkehard and Lisdat, Christian},
  journal={Metrologia},
  volume={58},
  number={1},
  pages={015005},
  year={2021},
  publisher={IOP Publishing}
}

@article{nakamura2025sub,
  title={Sub-femtosecond stabilization of multicore fiber for high-fidelity quantum networking at 100\% duty cycle},
  author={Nakamura, Takuma and Hoghooghi, Nazanin and Fontaine, Nicolas and Hayashi, Tetsuya and Nagashima, Takuji and Nardelli, Nicholas V and Reddy, Dileep V and Stevens, Martin J and Fortier, Tara and Shalm, Lynden K and others},
  journal={arXiv preprint arXiv:2509.17989},
  year={2025}
}

@article{nichol2022elementary,
  title={An elementary quantum network of entangled optical atomic clocks},
  author={Nichol, Bethan C and Srinivas, R and Nadlinger, DP and Drmota, P and Main, D and Araneda, G and Ballance, CJ and Lucas, DM},
  journal={Nature},
  volume={609},
  number={7928},
  pages={689--694},
  year={2022},
  publisher={Nature Publishing Group UK London}
}

@article{bersin2024development,
  title={Development of a {B}oston-area 50-km fiber quantum network testbed},
  author={Bersin, Eric and Grein, Matthew and Sutula, Madison and Murphy, Ryan and Huan, Yan Qi and Stevens, Mark and Suleymanzade, Aziza and Lee, Catherine and Riedinger, Ralf and Starling, David J and others},
  journal={Physical Review Applied},
  volume={21},
  number={1},
  pages={014024},
  year={2024},
  publisher={APS}
}

@article{kucera2024demonstration,
  title={Demonstration of quantum network protocols over a 14-km urban fiber link},
  author={Kucera, Stephan and Haen, Christian and Arensk{\"o}tter, Elena and Bauer, Tobias and Meiers, Jonas and Sch{\"a}fer, Marlon and Boland, Ross and Yahyapour, Milad and Lessing, Maurice and Holzwarth, Ronald and others},
  journal={npj Quantum Information},
  volume={10},
  number={1},
  pages={88},
  year={2024},
  publisher={Nature Publishing Group UK London}
}

@inproceedings{chung2021illinois,
  title={Illinois {E}xpress {Q}uantum {N}etwork ({IEQNET}): metropolitan-scale experimental quantum networking over deployed optical fiber},
  author={Chung, Joaquin and Kanter, Gregory and Lauk, Nikolai and Valivarthi, Raju and Wu, Wenji and Ceballos, Russell R and Pena, Cristi{\'a}n and Sinclair, Neil and Thomas, Jordan and Xie, Si and others},
  booktitle={Quantum information science, sensing, and computation XIII},
  volume={11726},
  pages={1172602},
  year={2021},
  organization={SPIE}
}

@article{kim2023improved,
  title={Improved interspecies optical clock comparisons through differential spectroscopy},
  author={Kim, May E and McGrew, William F and Nardelli, Nicholas V and Clements, Ethan R and Hassan, Youssef S and Zhang, Xiaogang and Valencia, Jose L and Leopardi, Holly and Hume, David B and Fortier, Tara M and others},
  journal={Nature Physics},
  volume={19},
  number={1},
  pages={25--29},
  year={2023},
  publisher={Nature Publishing Group UK London}
}

@article{clivati2020common,
  title={Common-clock very long baseline interferometry using a coherent optical fiber link},
  author={Clivati, Cecilia and Aiello, Roberto and Bianco, Giuseppe and Bortolotti, Claudio and De Natale, Paolo and Di Sarno, Valentina and Maddaloni, Pasquale and Maccaferri, Giuseppe and Mura, Alberto and Negusini, Monia and others},
  journal={Optica},
  volume={7},
  number={8},
  pages={1031--1037},
  year={2020},
  publisher={Optical Society of America}
}

@article{schioppo2022comparing,
  title={Comparing ultrastable lasers at $7\times10^{-17}$ fractional frequency instability through a 2220 km optical fibre network},
  author={Schioppo, Marco and Kronjaeger, Jochen and Silva, Alissa and Ilieva, R and Paterson, JW and Baynham, CFA and Bowden, William and Hill, IR and Hobson, Richard and Vianello, Alvise and others},
  journal={Nature communications},
  volume={13},
  number={1},
  pages={212},
  year={2022},
  publisher={Nature Publishing Group UK London}
}

@article{cantin2021accurate,
  title={An accurate and robust metrological network for coherent optical frequency dissemination},
  author={Cantin, Etienne and T{\o}nnes, Mads and Le Targat, Rodolphe and Amy-Klein, Anne and Lopez, Olivier and Pottie, Paul-Eric},
  journal={New Journal of Physics},
  volume={23},
  number={5},
  pages={053027},
  year={2021},
  publisher={IOP Publishing}
}

@article{clivati2022coherent,
  title={Coherent optical-fiber link across {I}taly and {F}rance},
  author={Clivati, C and Pizzocaro, M and Bertacco, EK and Condio, S and Costanzo, GA and Donadello, S and Goti, I and Gozzelino, M and Levi, F and Mura, A and others},
  journal={Physical Review Applied},
  volume={18},
  number={5},
  pages={054009},
  year={2022},
  publisher={APS}
}

@article{cizek2022coherent,
  title={Coherent fibre link for synchronization of delocalized atomic clocks},
  author={Cizek, Martin and Pravdova, Lenka and Minh Pham, Tuan and Lesundak, Adam and Hrabina, Jan and Lazar, Josef and Pronebner, Thomas and Aeikens, Elke and Premper, J{\"o}rg and Havlis, Ondrej and others},
  journal={Optics Express},
  volume={30},
  number={4},
  pages={5450--5464},
  year={2022},
  publisher={Optica Publishing Group}
}

@article{caldwell2025high,
  title={High-precision optical time and frequency transfer},
  author={Caldwell, Emily D and Triano, Theodora M and Sinclair, Laura C},
  journal={Advances in Optics and Photonics},
  volume={17},
  number={2},
  pages={375--440},
  year={2025},
  publisher={Optica Publishing Group}
}

@article{lindvall2025coordinated,
  title={Coordinated international comparisons between optical clocks connected via fiber and satellite links},
  author={Lindvall, Thomas and Pizzocaro, Marco and Godun, Rachel M and Abgrall, Michel and Akamatsu, Daisuke and Amy-Klein, Anne and Benkler, Erik and Bhatt, Nishant M and Calonico, Davide and Cantin, Etienne and others},
  journal={Optica},
  volume={12},
  number={6},
  pages={843--852},
  year={2025},
  publisher={Optica Publishing Group}
}

@article{li2024phase,
  title={Phase noise control of the fiber-based frequency dissemination system},
  author={Li, Wenlin and Chen, Yufeng and Dai, Hongfei and Wang, Fangmin and Wang, Bo},
  journal={IEEE Transactions on Instrumentation and Measurement},
  volume={73},
  pages={1--7},
  year={2024},
  publisher={IEEE}
}

@misc{bergquist1994laser,
  title={Laser stabilization to a single ion},
  author={Bergquist, JC and Itano, WM and Wineland, DJ},
  journal={Frontiers in laser spectroscopy},
  pages={359--376},
  year={1994},
  publisher={North Holland, Amsterdam}
}

@article{bercy2014,
  title = {Two-way optical frequency comparisons at $5\ifmmode\times\else\texttimes\fi{}{10}^{\ensuremath{-}21}$ relative stability over 100-km telecommunication network fibers},
  author = {Bercy, Anthony and Stefani, Fabio and Lopez, Olivier and Chardonnet, Christian and Pottie, Paul-Eric and Amy-Klein, Anne},
  journal = {Phys. Rev. A},
  volume = {90},
  issue = {6},
  pages = {061802},
  numpages = {5},
  year = {2014},
  month = {Dec},
  publisher = {American Physical Society},
  doi = {10.1103/PhysRevA.90.061802},
  url = {https://link.aps.org/doi/10.1103/PhysRevA.90.061802}
}

@article{duan2001long,
  title={Long-distance quantum communication with atomic ensembles and linear optics},
  author={Duan, L-M and Lukin, Mikhail D and Cirac, J Ignacio and Zoller, Peter},
  journal={Nature},
  volume={414},
  number={6862},
  pages={413--418},
  year={2001},
  publisher={Nature Publishing Group UK London}
}

@article{nardelli2025phase,
  title={Phase-Stable Optical Fiber Links for Quantum Network Protocols},
  author={Nardelli, Nicholas V and Reddy, Dileep V and Grayson, Michael and Sorensen, Daniel and Stevens, Martin J and Mazurek, Michael D and Shalm, L Krister and Fortier, Tara M},
  journal={arXiv preprint arXiv:2510.16230},
  year={2025}
}

@article{lindsey2019illuminating,
  title={Illuminating seafloor faults and ocean dynamics with dark fiber distributed acoustic sensing},
  author={Lindsey, Nathaniel J and Dawe, T Craig and Ajo-Franklin, Jonathan B},
  journal={Science},
  volume={366},
  number={6469},
  pages={1103--1107},
  year={2019},
  publisher={American Association for the Advancement of Science}
}

@article{liu2025urban,
  title={Urban sensing using existing fiber-optic networks},
  author={Liu, Jingxiao and Li, Haipeng and Noh, Hae Young and Santi, Paolo and Biondi, Biondo and Ratti, Carlo},
  journal={Nature Communications},
  volume={16},
  number={1},
  pages={3091},
  year={2025},
  publisher={Nature Publishing Group UK London}
}

@book{oppenheim1999discrete,
  title={Discrete-time signal processing},
  author={Oppenheim, Alan V. and Schafer, Ronald W. and Buck, John R.},
  year={1999},
  publisher={Prentice-Hall, Inc.},
  address={Upper Saddle River, NJ, USA}
}

@article{jabir2025enabling,
  title={Enabling phase stabilization of quantum networks via displacement-enhanced photon counting},
  author={Jabir, MV and Ahn, D and Annafianto, N Fajar R and Burenkov, IA and Battou, A and Polyakov, SV},
  journal={Optica},
  volume={12},
  number={5},
  pages={570--576},
  year={2025},
  publisher={Optica Publishing Group}
}

@article{newbury2007coherent,
  title={Coherent transfer of an optical carrier over 251 km},
  author={Newbury, Nathan R and Williams, Paul A and Swann, William C},
  journal={Optics letters},
  volume={32},
  number={21},
  pages={3056--3058},
  year={2007},
  publisher={Optical Society of America}
}

@article{liu2024creation,
  title={Creation of memory--memory entanglement in a metropolitan quantum network},
  author={Liu, Jian-Long and Luo, Xi-Yu and Yu, Yong and Wang, Chao-Yang and Wang, Bin and Hu, Yi and Li, Jun and Zheng, Ming-Yang and Yao, Bo and Yan, Zi and others},
  journal={Nature},
  volume={629},
  number={8012},
  pages={579--585},
  year={2024},
  publisher={Nature Publishing Group UK London}
}

@article{stolk2024metropolitan,
  title={Metropolitan-scale heralded entanglement of solid-state qubits},
  author={Stolk, Arian J and van der Enden, Kian L and Slater, Marie-Christine and te Raa-Derckx, Ingmar and Botma, Pieter and Van Rantwijk, Joris and Biemond, JJ Benjamin and Hagen, Ronald AJ and Herfst, Rodolf W and Koek, Wouter D and others},
  journal={Science advances},
  volume={10},
  number={44},
  pages={eadp6442},
  year={2024},
  publisher={American Association for the Advancement of Science}
}

\end{document}